**Frontiers of atom probe tomography physics, data processing, and analysis**

Emmanuelle A. Marquis[1,2], Arun Devaraj[3], Richard G. Forbes[4], Iman Ghamarian[5], Markus Kühbach[6], Jean-Baptiste Maillet[7], Baishakhi Mazumder[8], Frederick Meisenkothen[9], Jiayuwen Qi[10], Daniel Schreiber[2], Paul Styman[11], Francois Vupillot[7], Wolfgang Windl[10]

1 - Department of Materials Science and Engineering, University of Michigan, Ann Arbor, MI, U.S.A.

2 - Energy & Environment Directorate, Pacific Northwest National Laboratory, Richland, WA, U.S.A

3 - Physical and Computational Sciences Directorate, Pacific Northwest National Laboratory, Richland, WA, U.S.A

4 - School of Mathematics and Physics, University of Surrey, Guildford, Surrey GU2 7XH, UK

5 - School of Aerospace and Mechanical Engineering, The University of Oklahoma, Norman, OK, U.S.A.

6 - Physics Department and CSMB, Humboldt-Universität zu Berlin, Zum Großen Windkanal 2, D-12489 Berlin, Germany.

7 – University of Rouen Normandie, GPM UMR 6634, F-76000 Rouen, France

8 - Department of Materials Design and Innovation, University at Buffalo-SUNY, Buffalo, NY 14260, USA

9 - National Institute of Standards and Technology, Materials Measurement Science Division, Gaithersburg, MD, U.S.A., ORCID: 0000-0003-4045-0949

10 - Department of Materials Science and Engineering, The Ohio State University, Columbus, OH, U.S.A.

11 - United Kingdom National Nuclear Laboratory, D5, Culham Campus, Abingdon, Oxfordshire, OX14 3DB, UK

**Abstract**

Atom probe tomography (APT) fills a crucial need in the characterization workflow of materials by its ability to inform the 3D chemical microstructure at the nanoscale. As with any characterization techniques, APT has strengths and limitations that inform the interpretation of the data. Therefore, a challenge for the materials characterization community, and the APT community in particular, is the need to establish repeatable and reproducible workflows around the APT data acquisition, reconstruction, analysis, and sharing, in order to inform interpretation. Data interpretation also requires the continued development of our understanding of the physical processes responsible for field evaporation. We review recent developments in the experimental analysis of field evaporation and in the modeling of field evaporation leading to new understanding of common artifacts observed in reconstructed data. We then discuss current challenges with data analysis, translation of results, and data interpretation in the absence of community-agreed standards, and therefore, the crucial need for standardization at every stage of APT research, from data collection all the way to data reporting. This perspective is a summary of the invited presentations and discussions that took place during a workshop (August 4-5, 2024, Alexandria, Virginia).

# 1. Introduction

Atom probe tomography (APT) is a characterisation technique that holds great promise for atomic-scale materials characterisation, due to its unique combination of capabilities compared to other forms of mass spectrometry and tomography (e.g., theoretical equal sensitivity to all elements in the periodic table, site-specific sample preparation, sub-nanometer spatial resolution, 3D chemically resolved imaging, …). As a result, APT has become an essential technique in a variety of science and engineering fields. However, this "promise" can lead to expectations that are not always aligned with the current limitations of the technique.

These unrealistic expectations can erode confidence in APT as a measurement technique. Further, the scientific method is iterative and requires repeatability and reproducibility to function properly. Therefore, the workflow and parametrisation of APT sample preparation routines and the workflow applied for data processing, from calibration of the detector raw data to reported microstructural analysis, all must be repeatable and reproducible. This sentiment is echoed by publications (**Barker 2016**), which claim researchers in all fields, often cannot reproduce the measurement results of their peers – this is a problem. Without repeatability and reproducibility, APT researchers have a poor basis for decision making and scientific progress may be hindered.

APT data consists of millions of atoms extracted almost one by one from a specimen's surface through a process known as field evaporation (FEV). This physical mechanism was first identified in 1941 by E.W. Müller, who observed the desorption of barium atoms from a tungsten surface, leveraging the strong electric field generated when a needle-shaped specimen with a small apex is subjected to a high voltage (**Müller, 1941**). A positive electrostatic (ES) field, called here the *local surface field,* is induced by this voltage and (in conventional FEV theory) reduces the activation energy needed to allow the atom to escape from its bonding site. This leads to its ionisation and subsequent expulsion from the surface under the influence of the local surface field. This seemingly simple process forms the fundamental principle of APT, as conventionally understood. To perform time-of-flight mass spectrometry, field evaporation is carried out through a pulsed process. Control over FEV requires sub-nanosecond precision, meaning that voltage or thermal pulses must be applied within this timescale. To optimize analysis conditions, the FEV probability between pulses must be minimised. To prevent specimen failure and detector saturation, the evaporation rate or "flux" (measured in ions/pulse) is kept below 1 ion/pulse (in practice less than 0.01 ions/pulse). It is also crucial to maintain a homogeneous FEV flux density (as measured in ions per unit area per pulse) across the analyzed surface, as the assumption of regular evaporation is necessary to reconstruct accurate and well-calibrated atomic maps.

The primary experimental challenges in conducting APT characterization are:

1) <u>Mass spectrum analysis complications</u>: The APT community uses the term "ranging" as a name for the process of allocating mass-spectrum peaks to ionic chemical entities and selecting peak widths. APT's effectiveness in analyzing complex materials hinges on the precise assignment of peaks in the mass spectrum to chemical identities and in precise determination of peak widths. While the effective mass resolving power of current-day instruments (on the order to 500 to 2000) is sufficient for many applications, the process of ranging can face significant challenges for some materials, as exemplified by the case of oxides and nitrides. These materials commonly exhibit overlapping peaks within their mass spectra (e.g., $Mg++/C+$, $N+/Si++$, $N+$ with $N_2++$, and $O+$ with $O_2++$). This complicates accurate interpretation of the data. Additionally, peak width determination is both user-dependent and sensitive to experimental parameters (**Hudson, et al., 2011; Haley, et al. 2015**) with the presence of thermal tails that distort the mass spectrum. Deconvolution techniques based on natural isotopic abundances and peak form can mitigate some of these issues (**Meisenkothen, et al., 2020**), and optimizing experimental settings is critical to achieving increased mass resolution and improved signal-to-noise ratios (**Khan, 2016; Kelly, 2012**).

2) <u>Detection efficiency limitations</u>: Detection efficiency in APT remains a limitation for quantitative analysis, particularly when accurately measuring dopant concentrations, as not all ions reaching the detector are recorded. Signal loss can be attributed primarily to two factors: first, ions may arrive outside the timing window defined by voltage or laser pulses; second, "dead zones" within the detector's microchannel plate further hinder detection accuracy. Advances in microchannel plate technology have increased detection efficiency to around 52 % in reflectron-equipped systems and to 81 % in straight-flight configurations. However, reflectrons introduce a mesh that reduces detection efficiency, while significantly improving peak resolution (**Cerezo, et al., 1998**).

3) <u>Preferential ion loss</u>: The phenomenon of "preferential ion loss" has been reported in a wide range of materials, e.g., GaSb (**Muller, et al., 2011**), $Zn_4Sb_3$ alloys (**Gault, et al., 2010**), steel (**Caballero, et al., 2007**), semiconductors (**Ronsheim, et al., 2010**), and III-N structures (**Galtrey, et al., 2007**). Preferential ion loss refers to selective under-detection of a particular ion type. It is a particularly critical issue in III-N semiconductors and similar materials, where the under-detection of nitrogen is attributed to the in-flight generation of neutral fragments (resulting from molecular dissociation) that are not detected (**Di Russo, et al., 2018, Zanuttini, et al., 2017**). Another potential source of preferential ion loss are multi-hit events, referring to several atoms evaporating on the same pulse. Multi-hit events are influenced by both the bonding characteristics of the material and the species involved. For example, covalently bonded materials such as borides, carbides, and nitrides can exhibit higher incidences of multi-hit events due to strong atomic bonding and distinct field evaporation characteristics. Since current detectors have limited detectability when multiple ions arrive in close temporal and spatial proximity, multi-hit events necessitate advanced data correction methods to improve the reliability of quantitative APT measurements in these materials (**Martin, et al., 2017; Di Russo, et al., 2018**).

4) Spatial resolution constraints: Material heterogeneity (chemical via alloying and phases and structural via crystalline defects such as grain boundaries and interfaces) can influence spatial resolution, due to differences in the evaporation behaviour of atoms of different nature, atoms in different crystallographic or chemical environments. Spatial resolution also depends on the material being analysed. Unlike metals, oxide and nitride semiconductors typically require laser pulsing to induce field evaporation due to their strong covalent or ionic bonding. The spatial resolution in these materials is inherently limited compared to metals, due to differences in bond type and field penetration depth. While metals exhibit delocalised bonding and can "screen" the applied field, semiconductors' covalent bonds allow field penetration but restrict it to approximately 1 nm. This results in a less predictable evaporation pattern. Weakly bonded atoms evaporate randomly, while strongly bonded atoms exhibit correlated evaporation, further affecting spatial resolution (**Zanuttini, et al., 2017; Muller, et al., 2011**).

Given these limitations, one of the challenges for the APT community is to manage expectations and increase confidence in APT measurements. Broadly, we can do this by advancing the measurement science needed to assess APT accuracy, precision (repeatability), and reproducibility. Per the International Union of Pure and Applied Chemistry (**IUPAC, 2019**), accuracy is defined as closeness of agreement between a measurement result and the "true" value of the quantity that is subject to measurement, precision is the degree of agreement between independent test results when an experimental procedure is applied under specific conditions, and reproducibility is the degree to which independent results agree when using the same method on the same test material but under different conditions. Implicit in these three definitions is the need for standards. Accuracy cannot be measured without a reference value to compare against; precision cannot be evaluated without a defined experimental procedure;

and reproducibility cannot be assessed without a defined method. Currently, consensus-based APT standards do not exist, making it difficult to evaluate the accuracy, precision (repeatability), and reproducibility – this illustrates a clear need for APT standards development related to data evaluation.

The evaluation of APT repeatability and reproducibility necessitates the specification of methods and procedures, which inherently need to include data acquisition conditions. Unfortunately, it can be difficult to unambiguously define the actual APT acquisition conditions for a variety of reasons – e.g., differences in true sample temperature, laser alignment, true laser pulse fluence, geometry of the specimen being evaporated, etc. Since there is a direct relationship between the charge-state ratio (CSR) of the evaporated ions and the local surface field strength experienced by the sample apex, CSR can partially account for some of the differences between acquisition parameters that would otherwise be impossible to measure or control. However, CSR is not a perfect control metric, since chemical composition can vary throughout the sample and the emitted ion species and CSR often vary locally within the data point cloud.

Furthermore, APT data are complex and contains artifacts (e.g., reconstructed dimensions, ion trajectory aberrations, ion feedback signal, peak broadening, peak shifts, preferential ion signal loss, signal-to-noise/background variations, …). Many of these complexities are not yet fully understood and can be dependent upon the specific instrument used, or upon the specific software version used for data acquisition and/or data analysis. Analysts generally do not have access to the raw data collected by commercial instruments and must therefore work with data that has been influenced by "black box" algorithms that can significantly impact analysis results (**Meisenkothen, et al., 2020**), a situation that the instrument manufacturers could easily remediate.

The literature contains many examples of composition measurement bias, e.g., (**Morris, et al., 2018**), that are often field-dependent or depend on acquisition conditions. This has been attributed to a variety of reasons including, but not limited to, FEV behaviour (e.g., generation of neutrals, generation of compound ion species), multi-hit detection events, ion ranging methods, etc. Peak shapes can vary between different ion species, between spectra, and within spectra, leaving open the question of how best to assign regions of interest in the mass spectrum to obtain the most accurate and repeatable measurements. Due to the nature of APT data acquisition, it may not be possible to always find acquisition and analysis conditions that yield accurate analysis results. Therefore, an alternate approach to address composition measurement bias is needed. Reference materials, calibration curves, and standards-based analysis have been shown to be useful to improve analysis accuracy and precision. Standards-based analysis is an area of active research (**Smith, 1986; Meisenkothen, et al., 2020b; Gopon, 2022; DeRocher, et al., 2022**).

Accurate reconstructed length scales are also critical to achieving accurate quantitative microstructural and chemical analysis results. APT data often lack sufficient quality to resolve atomic planes in crystalline materials. Further, not all APT specimen materials are crystalline, or have crystallographic orientations that are known a priori. Therefore, the data point cloud provides no information that can be used as an internal length scale calibration for the reconstructed dimensions. Reconstruction algorithms have not yet developed to the point at which results on any samples can be implicitly trusted. Benchmark data and reference materials are, therefore, needed to evaluate, and support the development of, reconstruction algorithms.

The main challenges for the APT community are thus to increase our understanding of the mechanisms behind the generation of APT data, manage expectations, and increase confidence in APT measurements and data analysis. This demands a multi-prong approach. Closer examination of the physical processes describing the atomic escape is first discussed in **Section 2**. In parallel, due to the destructive nature of FEV and incomplete understanding of the underlying physics, forward modelling, which simulates the whole FEV process and tracks the trajectory of each atom (or molecule), can provide unique insights into atomic behaviour while allowing to quantify the uncertainties in reconstructed APT data with regard to the ground truth of the original emitter structure. Next, data analysis methods able to account for the spatial uncertainty and complex dependence with local chemistry and structure are needed, and recent attempts are described in **Section 3.** However, the black-box approach imposed by the instrument manufacturer undoubtedly constrains progress by limiting access to substantial amounts of information. We, therefore, conclude with a perspective on possible research avenues and with the APT community need for a profound cultural shift to adopt standardisation and openness that modern science requires.

## 2. Field evaporation: from mechanisms to forward modelling

APT data are reconstructed using models based on simplifying assumptions. Understanding where these assumptions may fail is a prerequisite to interpreting APT data. Reconstruction algorithms largely assume a hemispherical specimen shape where evaporation proceeds uniformly over the surface. In addition to relying on known materials properties, such as density and lattice constants, and on instrument geometry, scaling of the APT data relies on simple geometric relationships between the applied macroscopic voltage, $V$, the specimen apex radius, r, and the apex value of local surface field, $F_a$. This parameter $F_a$ is set equal to an appropriately defined "evaporation field". It was incorporated into the initial reconstruction algorithms and still plays a crucial role in subsequent updates (**Bas, et al., 1995; Larson, et al., 2013; Vurpillot, et al. 2013; Hatzoglou, et al., 2019; Gault, et al., 2021; 2008**). It is therefore crucial to understand how the term "evaporation field" is used and where some of the assumptions underlying the reconstruction algorithms might fail.

## 2.1. Current understanding of FEV

The original theories of FEV describe the escape process as thermodynamically determined, thermally activated, with only one escape charge-state involved (**Müller, 1956; Gomer, 1959**). The emission and detection fluxes (measured in ions/s, and also called "rates") are then assumed to be proportional to the evaporation rate-constant $K_{EV}$ (measured in inverse seconds) given by an Arrhenius-type equation:

$$K_{EV} = A exp\left(-\frac{Q(F)}{k_B T}\right) (1)$$

where $k_B T$ is the usual Boltzmann factor, $A$ is the field-evaporation pre-exponential, and $Q$ is the activation energy for field evaporation. In the simplest approximation, pre-factors are taken as slowly varying with $T$ and $F$, so that the evaporation rate-constant (and hence the evaporation rate/flux) is controlled by manipulating the applied voltage (and hence $F_a$) and/or the temperature, $T$, at the specimen surface. Later theories assumed that the atom might initially escape into a charge-state equal to 1, 2 or (in a few cases) 3, and might then be subject to field-induced post-ionisation. The activation energy, $Q$, is a microscopic parameter that brings together the various contributions that determine the energy required by an atom to be liberated, expelled, and detected as an ion, and FEV is a term that covers the whole process and mechanisms of escape and ionization. In the early ages of APT analysis, it was found that $Q$ was extremely sensitive to the applied electrostatic (ES) field. For metals, a variation of a few percent in the field can change the evaporation rate constant by several orders of magnitude (**Tsong, 1971**). It was also found that each element has its own magnitude of ES field required to cause significant FEV from the surface.

Empirically, $Q$ has been described via two simple properties: the FEV *operating field,* denoted here by $F_{op}$; and one or two parameters (described below) describing the "field- sensitivity" of $Q$, i.e., how $Q$ depends empirically on $F_{op}$. See (**Müller and Tsong, 1969**) or (**Biswas and Forbes, 1982**) for compiled lists of observed values of "operating" evaporation fields. Theoretically, for a given surface atom of a given element, the FEV activation energy barrier decreases to zero at a field now known as the *zero-barrier evaporation field* (ZBEF), denoted here by $F_{EV}$, (e.g., **Larson, et al., 2013; Miller and Forbes, 2014**). The operating field is always slightly or somewhat less than the ZBEF.

The ZBEF is one of the most fundamental parameters influencing data analysis. For a given element, a unique value of the ZBEF is often used. However, the local field at the specimen's surface is influenced by its geometric roughness at a given location, and the local ZBEF is also influenced by variations in local work function. The surface tends to form a stable end shape that adjusts to variations in evaporation rate-constant between different species present. Therefore, reconstruction artifacts can arise both from the dynamic morphological adaptations of the specimen, which is composed of different elements and/or phases with different ZBEFs on both the mesoscopic and microscopic scales, as shown in **Figure 1**, and from the inability of the present-day reconstruction algorithms to account for such deviations from ideality. Furthermore, at the microscopic scale, the presence of species with different ZBEFs can result in sporadic evaporation sequences. As the local surface morphology adapts to the sequential removal of atoms, retention effects and burst-evaporation mechanisms can stress the detector and result in the selective loss of atoms or a pile-up effect (**De Geuser, et al., 2007; Peng, et al., 2018; Russo, et al., 2018a; Meisenkothen, et al., 2015**).

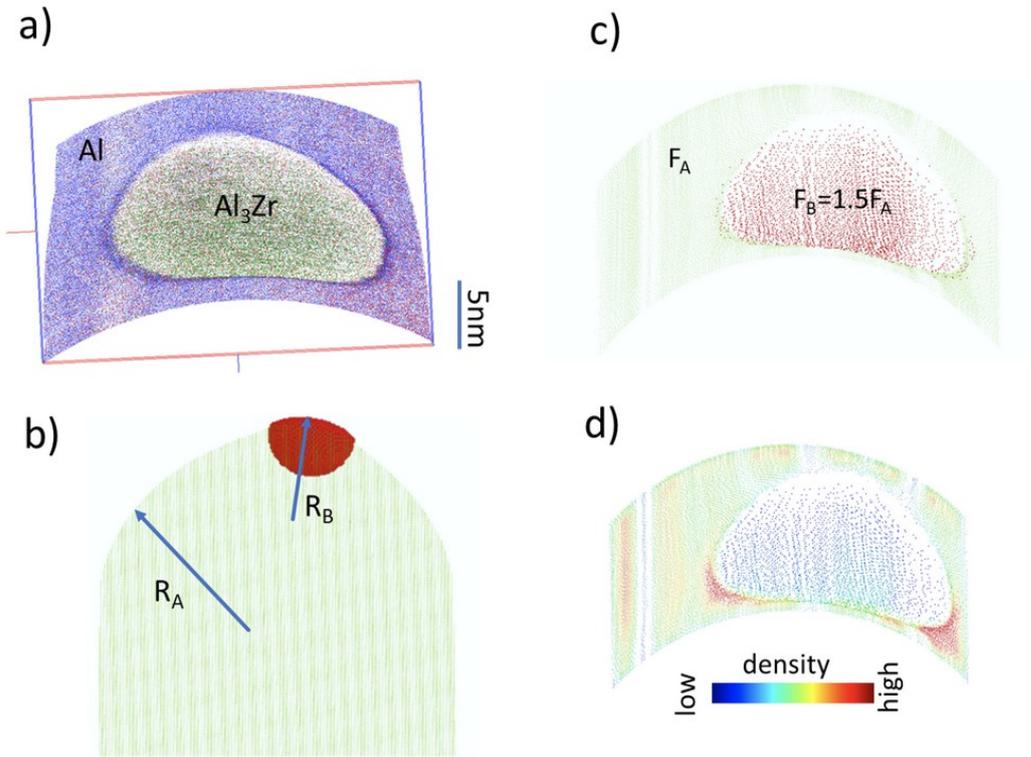

**Figure 1.** Example of the influence of evaporation fields on the reconstruction artifacts in APT. a) a spherical $Al_3Zr$ particle embedded in an Al matrix was analysed in APT. Some variations of the atomic density are observed b) Field Evaporation Models were used to reproduce the analysis. The $Al_3Zr$ is assumed to have an evaporation field 50 % stronger than the Al matrix. It generates dynamic evolution of the curvature during the analysis c) and d) the particle is reconstructed with similar reconstruction artifacts (local magnification effects and trajectory overlap at the interfaces between phases)

Experimentally, local variations in the evaporation behavior can be assessed and understood via the field sensitivity, $S_T(T,F)$, of the evaporation rate-constant close to the ZBEF. As defined by Brandon (**Brandon, 1966; McKinstry, 1972**), this sensitivity can be defined mathematically by: $S_T(T,F) = \left(\frac{dln\,K_{EV}}{dlnF}\right)_T = -\frac{F}{k_BT}\left(\frac{dQ}{dF}\right)$. Close to $F_{EV}$, we can define the quantity $C = F_{EV}\frac{dQ}{dF}$, so that $C \sim k_BT \cdot S_T(T,F_{EV})$. A simple assumption takes the activation energy barrier for the evaporation of the considered atom to be given adequately by the linear approximation:

$$Q(F) = C\left(1 - \frac{F}{F_{EV}}\right) \qquad\qquad (2).$$

Non-linearities of $Q$ linked to the complexity of the field evaporation mechanism exist (**Waugh, et al., 1976; Forbes, et al., 1982; Wada, 1984; Miller & Forbes, 2014; Ashton, et al., 2020; Vurpillot, et al., 2024**). However, when they can be considered a small effect in the regime used experimentally (typically $F_{op}$ in the range 0.8 $F_{EV}$ to 0.95 $F_{EV}$), then the parameter $C$ can be treated as constant, and therefore field evaporation can be investigated experimentally.

The stronger the $C$-value for a given element, the more deterministic the evaporation process can be. In alloys, preferential FEV of species can be rationalized by assuming that differences in $C$-value between species are large (**Hatzoglou, et al., 2020**). Therefore, the $C$-value can be used to understand evaporation behaviour during a voltage-pulsed experiment. For example, the shape of the mass peaks obtained during voltage pulsed analyses can be understood through variations in the $C$-value (**Rousseau, et al., 2023; Vurpillot, et al., 2024**). Indeed, ions are assumed to leave the surface at the top of the pulse, acquiring almost instantaneously all the kinetic energy induced by the potential energy. Due to the probabilistic nature of the FEV process (**Eq.1**), some ions can leave the specimen slightly before or after the top of the pulse, at a voltage $V = V_T(1 - \delta)$, where $V_T$ is the maximum applied voltage from the combination of DC and pulsed high voltages. Hence, ions can have an energy deficit, $\delta$, leading to an increase of their time-of-flight, which creates a tail in the observed mass spectrum. An illustration of this process is provided in **Figure 2**, where a statistically significant number of atoms from a pure Fe (iron) specimen was evaporated in voltage pulse mode enabling the mapping of the $C$ constant over the detector (**Figure 3b**). This analysis highlights the crystallographic dependence of FEV. Indeed, the field sensitivity of the evaporation process is dependent on the path followed by the departing atoms during the FEV process, which depends on local bonding.

However, the parameter $C$ needs to be evaluated theoretically, and the correct interpretations developed. To this end, the recent use of ab initio methods that consider the dynamic motion of the atom in the first step of field evaporation shows a great leap forward compared to both the older scientific models that culminated in the so-called "Revised Charge Draining (RCD) mechanism" (**Forbes and Chibane, 1986**) and the phenomenological **Eq. (2)**. Neither have been able to predict the precise form of the $Q$ equation or precise values for $C$. In the energetic interpretation of FEV, density function theory (DFT) models have also confirmed the significant influence of the initial trajectory in the first stage of evaporation (now considered to normally be so-called "roll-up motion"). This has long been suspected as a possibility, e.g., (**Waugh, et al., 1975; Forbes, 1995**). Modern DFT calculations (**Ashton, et al., 2020; Ohnuma, 2021**), and the new technique of *Field Evaporation Energy Loss Spectroscopy* (FEELS) (**Vurpillot, et al., 2024**), have confirmed this as usually likely, and we can now look forward to quantitative predictions of field-sensitivity parameters.

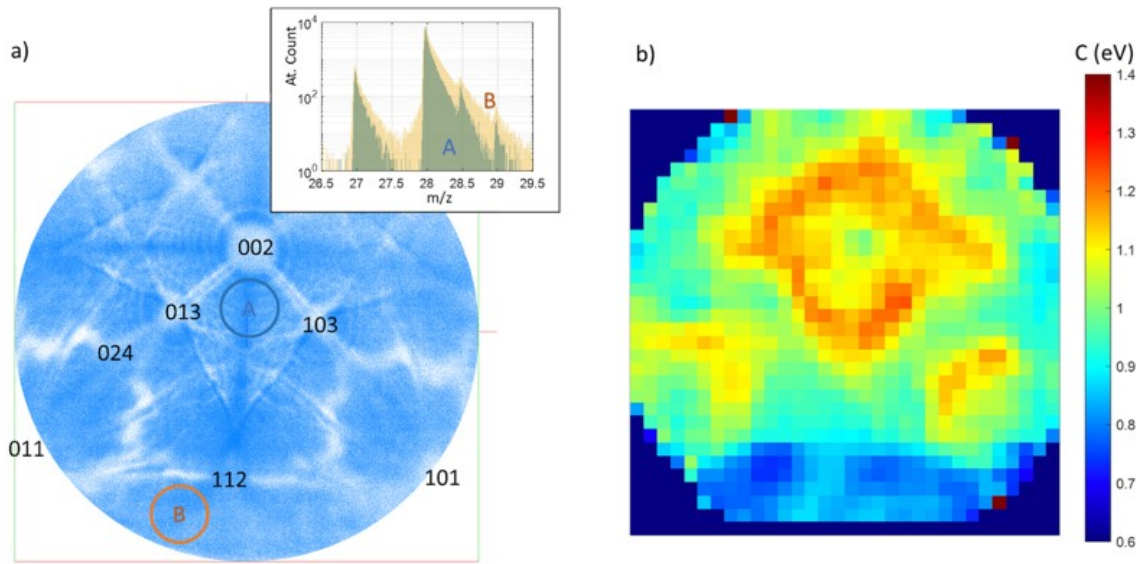

**Figure 2**: a) impact map of Fe2+ onto the detector showing density fluctuations induced by the crystallography of the specimen (BCC structure), evaporated using a LEAP5000 in voltage mode at 50 K. Some crystallographic directions are indicated on the map. Mass spectra of regions A and B are reported in inset showing the change in tail slopes between regions. b) Computation of the C map from the mass tails, showing the strong influence of crystallography on this parameter (C varying here between 0.7 eV and 1.3 eV)

The RCD FEV mechanism and the modern phenomenological approach depend on knowledge of the electric field value at the surface. This is a determining variable that influences composition analysis (**Ndiaye, et al., 2023; Rigutti, et al., 2016; Mancini, et al., 2014**), multiple events (**Russo, et al., 2018**), surface diffusion (**Marquis, et al., 2008; Oberdorfer, et al., 2018**) and 3D reconstruction (**Xu, et al., 2015; Hatzoglou, et al., 2017; Marquis, et al., 2011**). However, this variable remains difficult to establish during an APT analysis. It can nevertheless be estimated by applying post-ionization theory directly to the data, using the so-called Kingham's curves (**Kingham, 1982**). These curves, although probably not exactly correct, provide a useful relationship between the charge-state ratio (CSR) and the surface electric field during evaporation. Note that a mathematical procedure for inverting the relationship, to derive the surface field from the CSR, was recently proposed (**Tegg, et al., 2024**). An averaged CSR can be computed locally for each species of evaporated atom, from which an averaged local surface electric field is inferred.

In **Figure 3**, this method is applied to an APT volume obtained from an Al-Sc alloy containing $Al_3Sc$ precipitates. The only tabulation of estimated ZBEF values that includes both Al and Sc appears to be that of (**Miller and Forbes, 2014**), which gives values of 18 V/nm and 29 V/nm, respectively. For Al, theoretical ZBEF estimates of 19 V/nm or thereabouts are given by several other sources, e.g., (**Tsong 1990; Sanchez, et al. 2004; Tegg, et al., 2024**). There are no known estimates of operating fields for elemental Sc and the published operating-field estimates for elemental Al do not make sense; (**Boyes, et al., 1977**) reported a value of 27 V/nm, and (**Tsong, 1990**) reported a value of 33 V/nm. In practice, the observed presence of Al2+ ions indicates a higher operating-field during the FEV of this phase, probably between 20.6 V/nm and 21 V/nm, which would suggest a variation difference of ~1 V/nm between matrix and precipitates, indicating corresponding to an operating-field variation around 5 %.

As reported in the literature, even a a variation as small as 5 % can affect the reconstruction of precipitates (**De Geuser, et al., 2007**). This occurs as a result of trajectory aberrations due to local distortions of the surface specimen morphology (the local magnification effect): these may also alter the spatial resolution and their measured composition. This morphological modification is directly observed here within the analysed APT volume. Precipitates that (according to TEM imaging) are expected to be spherical appear elongated into oval shapes after reconstruction.

This new approach for determining the local evaporation field using the nearest-neighbour method represents a potential improvement in addressing the biases introduced by certain local field variations that remain almost indistinguishable with current data treatment. It could thus contribute to a better understanding of atom probe data and encourage caution in their interpretation.

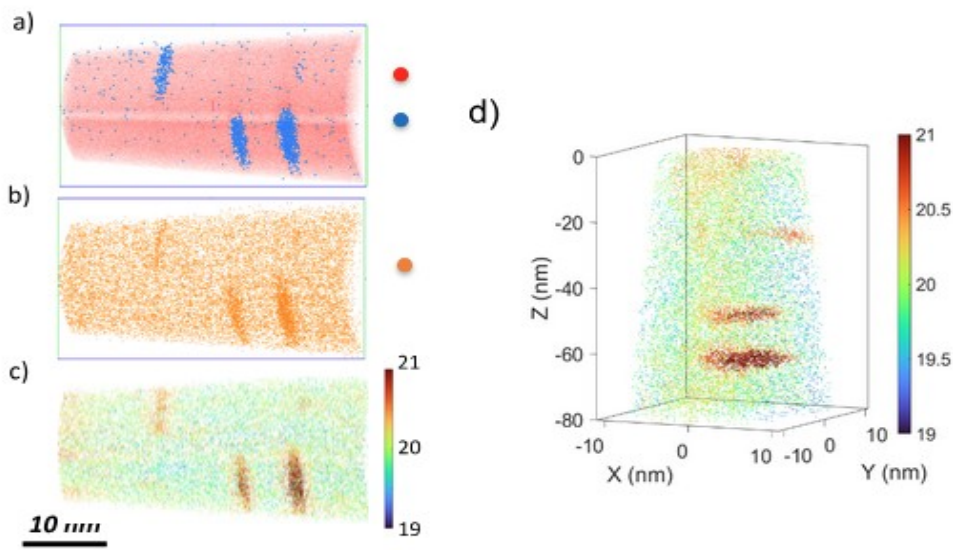

**Figure 3:** (a), (b) Represent an APT volume of an Al sample with Sc precipitates (c) and (d) the same volume obtained with the local evaporation field estimation LEFE represented by color for each Al²⁺ atom.

## 2.2. Current forward-modelling approaches

Parallel to the analysis of the data to inform on physical mechanisms, forward-modelling approaches provide insights into the FEV behaviour of complex materials and add another layer of interpretation for the APT data. The foundational scheme was introduced by Vurpillot (**Vurpillot, et al., 1999**). In the first iterations, following Arrhenius-type theory (**Eq. 1**), the atom with the lowest activation energy and thus the highest evaporation rate-constant, i.e., highest FEV probability per unit time, was selected to evaporate: its trajectory was classically integrated using Newton's equations of motion (**Vurpillot, et al., 2015**). Therefore, the approach relied on the input of activation energies or energy barriers against field evaporation to select the evaporating atom, which normally depend on the atom's chemical nature, its environment, and on its response to the electric field.

Although it has never been explicitly stated in the APT literature, the proper classical theoretical basis for the thermal-activation-based escape models is the well-established harmonic transition state theory (hTST) (**Eyring, 1935; Vineyard, 1957**). When examining the FEV process in the light of the basic assumptions underlying hTST, especially in the context of practical experimental conditions, several concerns arise. First, it is assumed that the system is in thermal equilibrium. However, this is not the case in experiments. During voltage-pulsed evaporation, the applied voltage varies with time, leading to continuous changes in the electric field and the surface-atom physical environment. At any given instantaneous value of applied voltage, this causes the system to deviate from the thermal equilibrium situation corresponding to this applied voltage. To account for the time dependence of the evaporation rate-constant from the pulse contribution, a correction to the activation energy has been previously suggested (**Vurpillot, et al., 2016**). Another possibility would be to interpret the parameter $T$ as an "effective temperature", as is done with semiconductor FEV.

During laser-assisted FEV of semiconductors, the picosecond response time of field evaporation is much slower than the femtosecond pulse length of the laser used in most current instruments (which pushes the whole emitter far from thermal equilibrium). The issues are how energy is absorbed from the laser pulse and transferred (ultimately) to the evaporating atom, and what the times involved are. A hot-phonon-assisted mechanism was proposed to explain this (**Silaeva, et al., 2015**). It was also suggested that the temperature parameter in the Arrhenius equation should be replaced by an "effective phonon temperature" for the surface, to account for the thermal effect induced by the laser, which lasts much longer (picoseconds) than the ultrafast laser-pulse duration (femtoseconds) (**Silaeva, et al., 2015**).

Another limitation of basic activation-based escape models is the use of the classical image potential energy (PE) to describe the interaction between the ionised adatom and the substrate. This PE is derived for a system where a point charge is placed above an infinite metal surface and is well known to be an approximation with limited accuracy. For a fuller discussion, including that of other PE terms involved in fuller classical or quantum theories, see Section 3.2.8 in (**Miller and Forbes, 2014**). By calculating energy barriers using DFT (**Yao, et al., 2015; Ashton, et al., 2020**), it was shown that the relationship between barriers and surface fields differs significantly for atoms departing from a kink site compared to an adatom site. This confirms decisively that it is not a good approximation to replace actual interatomic forces with image charge forces when modelling field evaporation of atoms on the emitter surface. These DFT calculations also reproduced the known experimental result that FEV preferentially takes place at kink sites (because these sites have the fewest nearest neighbours and therefore are bonded weakest).

Considering all the above, is there an alternative approach to describe evaporation events in forward modeling? FEV can be classified as an "infrequent event" in the sense that the average time between evaporation events is very large compared to vibrational periods (~picoseconds). Over the past 30 years or so, (accelerated) molecular dynamics (MD) simulations have been proposed as an alternative for "complex infrequent event systems" which are outside of the validity range of hTST (**Voter, 2012**). The classical equations of motion are integrated over time to generate ion trajectories, regardless of whether the system is in thermal equilibrium or not. The interatomic forces can be determined either in an ab initio way by an electronic structure method such as DFT, often labelled as ab initio MD (AIMD) (**Car, 1985**), or by classical MD which employs interatomic potential functions to represent force fields. While AIMD is more accurate, it is computationally expensive and therefore not feasible for large systems on longer time scales without significant advancements in computer power and algorithm efficiency. In contrast, classical MD is more computationally efficient but is limited in accuracy only by chosen interatomic potentials, making it well-suited for modelling FEV, which involves systems with millions of atoms. Therefore, the MD approach to complex infrequent event systems was integrated into the so-called TAPSim-MD method (**Qi, et al., 2022**), where FEV is integrated as part of an MD simulation. There, the forces on the atoms are approximated by the superposition of the zero-field interatomic forces from traditional MD and the field-induced electrostatic forces derived from the solution of Poisson's equation. This approach can capture key features of FEV, allowing evaporation events to occur naturally during the MD simulation, at the critical evaporation field above which the induced electrostatic forces exceed the interatomic bonding forces.

The TAPSim-MD's dynamic approach has allowed for a detailed investigation of the bond-breaking processes that lead to evaporation events. For instance, in [001]-oriented γ-TiAl, which consists of alternating Ti and Al atomic layers, most APT experiments find that Ti atoms evaporate quickly following the evaporation of Al in the layer above, suggesting that Ti has a lower evaporation field than Al. This conclusion seems counterintuitive considering that Ti has much stronger bonds than Al in the intermetallic phase. However, as shown in **Figure 4(a)**, TAPSim-MD simulations reveal distinctly different trajectories for evaporating Ti and Al ions, where Ti breaks its bonds in two steps, while Al breaks all its bonds at once (**Qi, et al., 2025**). This two-step bond-breaking process explains why Ti, despite its stronger bonding, evaporates more easily—the force required to break one half of the bonds at a time in two steps is less than the force needed for Al to break all its bonds at the same time.

The full dynamics in TAPSim-MD also enables more realistic integration of ion trajectories. In the traditional modelling scheme, since no dynamics are used to select evaporation events, an initial velocity of zero is assigned to the escaping ion, and its trajectory is determined solely by the electrostatic force in the applied field. This approach thus fully neglects the contribution from interatomic interactions during and after bond breaking. As a result, the traditional modeling approach is unable to capture any dynamic effects in APT data (**Qi, et al., 2023**). In contrast, MD involves the full dynamics of field evaporation, providing a more realistic simulation of ion behaviour and trajectories for predicting detector maps and reconstruction artifacts. An example of the importance of considering such dynamic effects is the artifact of enhanced zone lines observed on detector maps. Analysis of the atom trajectories shows that during field evaporation, some ions depart from the surface with velocities whose directions are not aligned with those of the local fields. This deviation arises from the misalignment of the electrostatic force, which is normal to the envelope of the local surface, and the interatomic force, which is anisotropic and depends on the configuration of the neighbour atoms. For the canonical example of tungsten, **Figure 4(b)** shows deviations in the lateral velocity components of atoms relative to their local electric fields, which are most pronounced in the vicinity of the <111> and <100> zone lines. These deviations point towards the zone lines, leading to the enhanced intensity as observed in experimental detector impact maps.

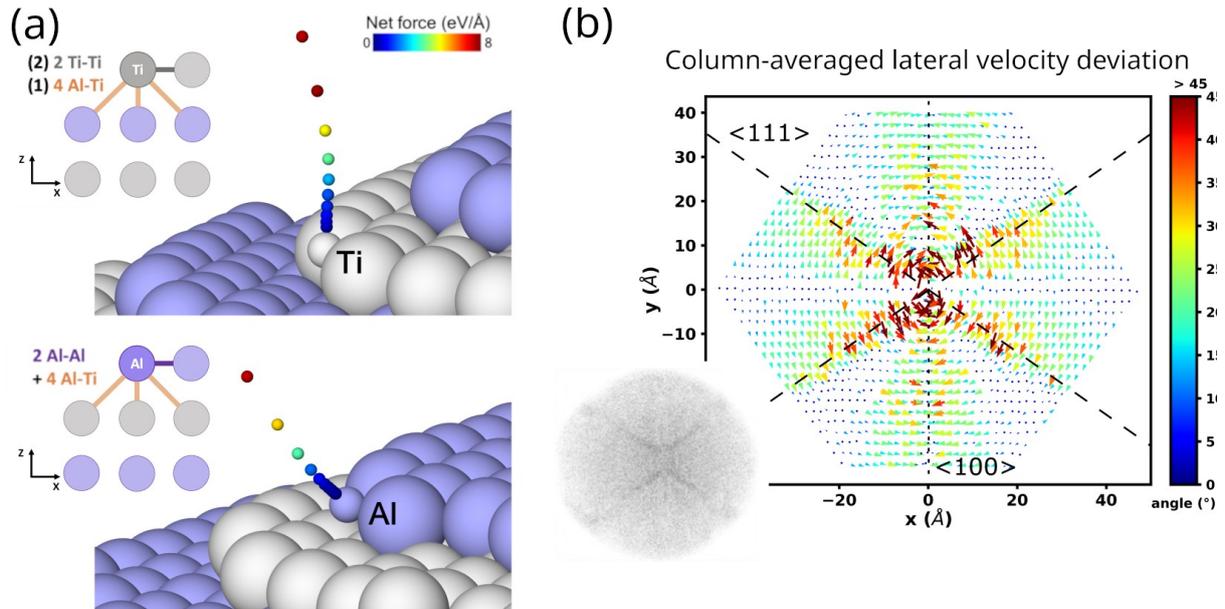

**Figure 4**: (a) Examples of Ti (grey) and Al (violet) atom trajectories during field evaporation, accompanied by schematic diagrams illustrating their bonding conditions and bond-breaking sequences respectively. The atom trajectories are described by a series of small dots colored by the magnitudes of net forces at different instances. (b) Column-averaged lateral component of launch velocity in the xy-plane, with an inset of the experimental field evaporation map for a ⟨110⟩-oriented W tip, reprinted from (**Qi, et al., 2023**). Vectors are coloured by the angle between launch velocity and the external force. Surface atoms are coloured by the magnitude of the electric-field-induced forces.

Looking ahead, while electrostatic-force-coupled MD is currently the most computationally economical approach for simulating APT emitters containing millions of atoms. Its accuracy, however, is constrained by the fidelity of the interatomic potentials and by how surface charges are computed. Moving towards a more ab initio description will ultimately require fully quantum-mechanical AIMD, which may become practical for APT systems as exascale computing and first-principles algorithms continue to advance. In addition, despite the successes of the electrostatic-force-coupled MD, several apparent scientific difficulties with the "bond breaking first" (BBF) mechanism have been pointed out informally (**Forbes, 2024**). These include: (a) significant discrepancies between BBF critical evaporation fields ($F_c$) and reported *observed* operating fields ($F_{op}$) (for Cu, Ni and W, the ratios $F_{op}/F_c$ are approximately 2.2, 2.5, and 3.0, respectively); (b) apparent inability of the BBF mechanism, in its present form, to provide a good explanation of the observed temperature dependence of operating field, e.g., (**Vurpillot, et al., 2006**); and thus (c) apparent inability of the BBF mechanism, in its present form, to explain clearly how the laser-pulsed atom probe works; and (d) apparent inability of the BBF mechanism, in its present form, to explain the experimental existence of what has hitherto been interpreted as a tunneling regime that operates at sufficiently low temperatures (**Menand and Blavette, 1986**). Further research is needed to resolve these issues.

## 3. Best practices

### 3.1 Advances in data analysis

It can sometimes by challenging to determine whether observed patterns in solute distributions within APT data are meaningful representations of material microstructures or simply random fluctuations. This uncertainty can undermine the confidence in conclusions drawn from APT studies, particularly in applications where precise characterisation of nanoscale features is crucial. For example, the absence of a reliable method to quantify the statistical significance of spatial distributions in APT data presents a significant obstacle to the field's advancement. The issues arise from multiple sources. As highlighted above, field evaporation artifacts compounded by limitations in the reconstruction algorithms can generate significant uncertainty in the positioning of atoms within the reconstructed volumes. These challenges become particularly acute for the analysis of small solute clusters or solute clouds within APT datasets.

Traditional quantification methods for solute clustering, such as the maximum separation method (MSM) and its variants, have been widely used but often struggle with datasets where the composition difference between clusters and matrix is not pronounced (**Hyde & English, 2011; Marquis & Hyde, 2010**). The output of the methods are also questioned due to the lack of objective approach guiding the selection of input parameters, which often rely on user expertise and can lead to subjective results. The lack of reporting of the selected values of the input parameters and even the algorithm used is also troubling. Furthermore, there is a limited understanding of the spatial statistics inherent in APT data, particularly in determining whether deviations from complete spatial randomness are statistically significant or merely artifacts of the data collection process. These issues are exacerbated by the fact that many computational algorithms do not scale efficiently and therefore analyzing large point cloud becomes costly. This restricts the ability to perform comprehensive statistical evaluations or iterative optimisations. To try to address some of these limitations, a hierarchical density-based cluster analysis (CHD) method (**Ghamarian, et al., 2019**), combining the HDBSCAN and DeBaCl algorithms, offers improved cluster detection independent of morphology and can handle clusters of varying densities within the same dataset. This method also provides quantitative measures of clustering performance through stability scores and atom-level probability assignments.

Beyond detection, the automated classification of the detected microstructural features is also needed. For example, the analysis of dislocations has also been done manually (**Williams, et al., 2011**). A semi-automated approach combining dense object identification, skeleton-finding algorithms, and object classification was proposed to discern and analyse dislocations in APT datasets (**Ghamarian, et al., 2020; Saxena, et al., 2024**). **Figure 5** demonstrates the complex effects of irradiation on the spatial distribution of Si solutes. The irradiation process induces the formation of condensed, ellipsoidal Si-rich precipitates while simultaneously promoting the segregation of Si atoms to dislocations. This figure showcases the efficacy of the methodology in detecting Si-rich dense objects and reliably categorizing them based on their distinct morphological characteristics.

Several promising research directions are currently emerging, continuing the transition from deterministic cluster analysis methods to probabilistic approaches that can account for uncertainty in the data and in the analysis process. Instead of reporting a single, definitive number of clusters, these methods would generate a probability density function for the number of clusters in an APT dataset. Such probabilistic representation would enable more nuanced comparisons between datasets using metrics such as Kullback-Leibler divergence, providing a more robust foundation for statistical analysis in comparison to the currently used deterministic cluster analysis methods. Additionally, there is a pressing need to develop computationally efficient tools to perform advanced statistical characterisation of APT datasets using sophisticated summary function methods. These tools would improve the accuracy and reliability of cluster analysis and open new avenues for exploring the complex spatial relationships within APT data.

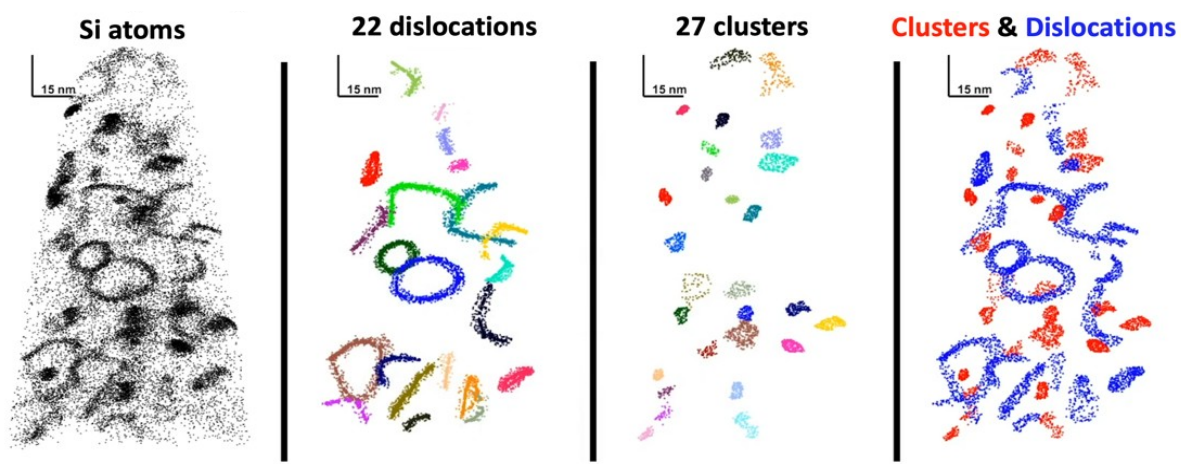

**Figure 5.** Spatial distribution and morphological classification of Si-rich features in an irradiated APT sample. From (**Ghamarian, et al., 2020b**). Left image -- point cloud showing the reconstructed silicon atom positions in the sample. Left Center image -- each color indicates the silicon atoms associated with a specific dislocation. Right Center image -- each color indicates the silicon atoms associated with a specific local cluster of silicon atoms. Right image -- overlay of dislocation image with cluster image. Blue points indicate silicon atoms associated with dislocations and red points indicate silicon atoms associated with clusters.

Other advanced data analytics methods are emerging, providing sophisticated insights into the structural and chemical complexities of materials, such as nitride and oxide semiconductors. For example, Licata, et al. addressed the substantial challenges posed by field evaporation inconsistencies when examining dopant-defect interactions in Mg-doped GaN using APT (**Licata, et al., 2021**). These inconsistencies often hinder accurate quantification of dopant distributions. Their approach incorporates principal component analysis (PCA) and voxelisation, which together mitigate artifacts from non-uniform field evaporation, facilitating a reliable investigation of Mg-rich clusters and dislocations. Recent research has also demonstrated the efficacy of combining APT with unsupervised machine learning (ML) modelling to study phase transformations in complex alloy systems, such as $(Al_xGa_{1-x})_2O_3$ films (**Sarker, et al., 2020**). These films exhibit notable structural transformations as the Al content varies from 10 % to 100 %. By integrating PCA with APT data, subtle yet meaningful shifts in alloy composition were detected, marking phase transformations and underscoring data-driven methods' capability to extract hidden patterns within extensive datasets. These findings were corroborated using complementary techniques such as scanning transmission electron microscopy (STEM) and X-ray diffraction (XRD), validating the structural phase transitions observed with APT (**Sarker, et al., 2020; Bhuiyan, et al., 2020**). More generally, ML approaches are fast being proposed to develop automated APT workflows for mass ranging (**Meisenkothen, et al., 2020; NIST Statistical software, 2025**), crystallographic analysis, fine precipitates detection, and chemical topologies in a range of materials (**Li, et al., 2021; Saxena, et al., 2023; Li, et al., 2026**).

### 3.2. Collecting and analysing data and metadata

APT experiments require user input, and more importantly, user choice at every stage of the experiment – be it from sample preparation, collecting the raw data, calibrating these data into reconstruction of the 2D data into 3D volumes, spectrum peak assignment and peak-width selection (hereafter called mass ranging), data analysis, and reporting, as outlined in **Figure 6**.

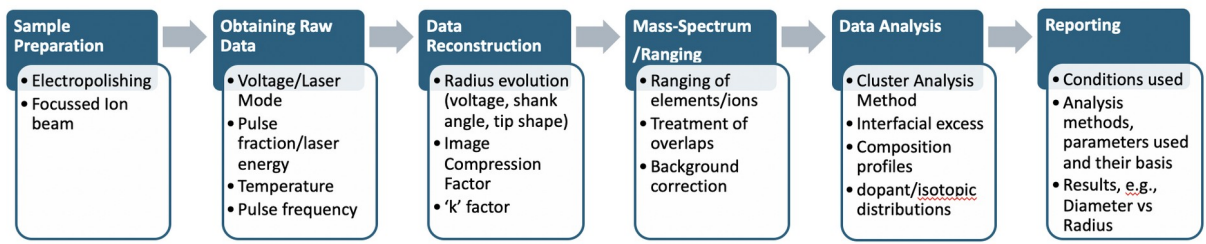

**Figure 6**: Stages of an APT experiment and some of the choices made by users.

The user-defined choices influence the quality of the data and affect subsequent quantitative analysis. Poor choices of experimental parameters mean that even the most diligent and careful analysis will be performed on data that may be of low quality, compromising the results. For example, setting too low a pulse fraction may lead to preferential loss of particular ions between pulses, which would then be systematically underestimated in the subsequent analysis (**Hatzoglou, et al., 2020; Hyde, et al., 2011**). Similarly, high quality raw data can lead to inaccurate results if subsequent data reconstruction and analysis are performed poorly.

It is perhaps easier to make an assessment of the impact of data acquisition parameters on data quality since users can compare measured sample compositions to known compositions, or let mass-spectra provide a visual indication of high background or large peak tails. Therefore, selecting optimal conditions for their materials is measurable and less subjective, as demonstrated in the case of reactor pressure vessel steels by Hyde, et al. (**Hyde, et al., 2011**).

However, in the case of data analysis (and to some degree mass-ranging), choices can be more subjective since in this case the 'exact' or 'correct' answer is unknown. Some guidelines and tools are available to support mass-ranging and treatment of peak overlaps (**Haley, 2015; Hudson, 2011; Mikhalychev, 2020; London, 2017; Wei, et al., 2025**). Additional methods such as automated elbow-ranging as described in (**Grimm, et al., 2025**) could be considered to improve ranging. There are multiple algorithms for performing cluster analyses on APT data, all requiring selection of parameters which have a large influence on the results. The most commonly applied is the maximum separation method (**Hyde 2011**), but many other methods exist, such as isoposition (**Lefebvre-Ulrikson, 2016**), and there are many papers suggesting optimal methods to select parameters, which were summarized in (**Dong, et al., 2019**).

Several benchmarking studies have shown that the differing analysis methods as well as user-to-user variation due to parameter selection can be a significant source of uncertainty/variability in the results (**Dong, et al., 2019; Marquis, et al., 2018; Hyde, et al., 2017**). This user-to-user variability can result in difficulties when collating data from multiple sources to compare with modelling. Castin, et al. also showed the variability in analysis of RPV steels led to difficulties in extracting trends of cluster size and number density with irradiation dose (**Figure 7**). When consistent analyses were performed, trends became easier to extract and consistent with model predictions (**Castin, et al., 2025**).

Therefore, there is a clear need for consistency/best practices in particular for data analysis, but also for data acquisition. Recommendations of optimal conditions will need to be material specific, but overarching guiding principles on how to select conditions may be useful. For data analysis, it is increasingly clear that, for example, single values of cluster size do not portray the complete picture, and thus the size is better represented by the distribution. Best practice guidelines would be very beneficial for the reporting of APT data and could build upon the recommendations by Blum, et al. (**Blum, et al., 2017**) and Dong, et al. (**Dong, et al., 2019**), but could be expanded to include treatment of peak overlaps and consistency in reporting of numbers, e.g., there are multiple descriptions of radius (equivalent sphere, Guinier, gyration). This would enable better understanding between researchers as to how the data have been acquired, reconstructed, ranged, and analysed. Sharing of data, complete metadata, and standardisation ought to be embraced by the APT community.

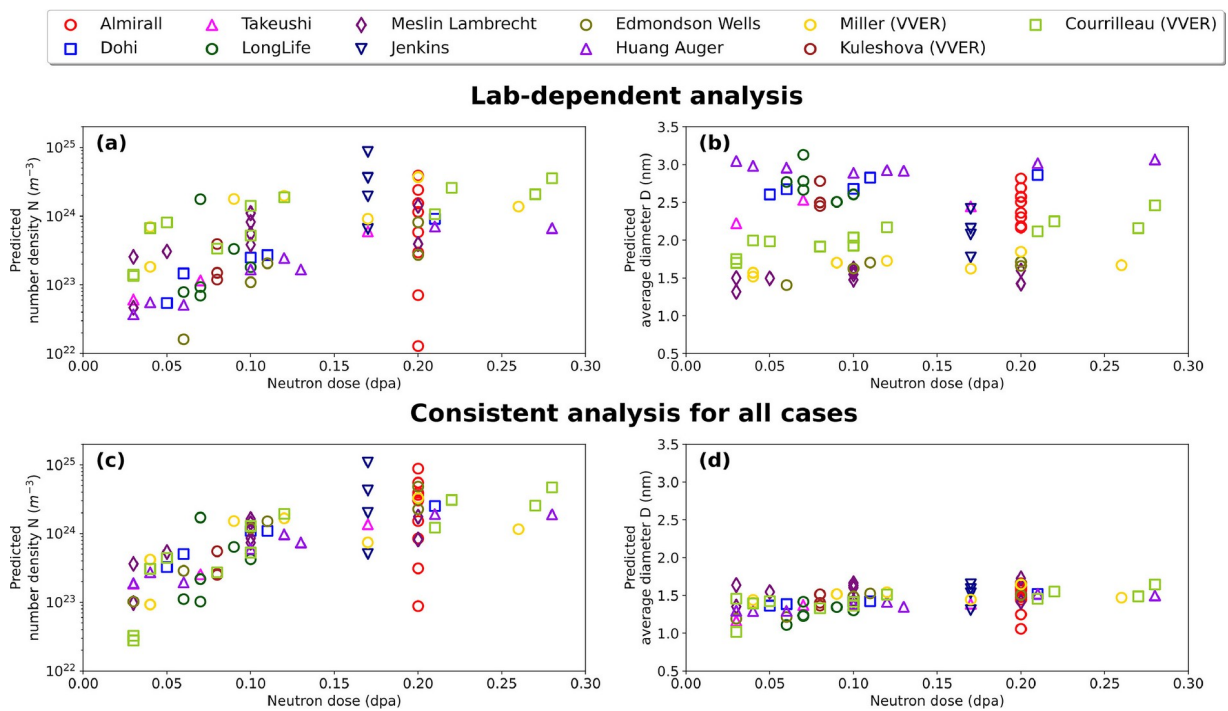

**Figure 7:** Comparison of cluster size and number density predicted by OKMC model when analysed using parameters determined by different labs (top) and consistently by a single set of parameters for all datasets (bottom). From (**Castin, et al., 2025**).

### 3.3. Sharing data and metadata

Materials characterisation with APT involves integrating experimental methods (**Kühbach, et al., 2022**) and computational techniques (such as data analysis and modelling) into structured workflows (**Kühbach, et al.**, **2024**). The collective knowledge in this field is built from the expertise of several hundred specialists, publications, and the tools developed with the leading industry partner. Knowledge exchange in APT research predominantly occurs through publications, conference presentations, and collaboration among research groups. Currently, most of the APT literature remains behind paywalls or is inaccessible due to poor metadata. For example, a survey of articles in ScienceDirect with "atom probe" in the title, abstract, or author-specified keyword done in 2024 revealed that only 365 out of the total 3,914 were open access, and just 76 offered data upon request. Barriers include a conflict of interest with data still used in ongoing research, technological limitations, and lack of time. The few APT data that are shared publicly via platforms such as Zenodo, include reconstructed data, but often lacking essential ranging files or comprehensive, automated metadata.

This reality is in contrast to the expectations that in today's digital environment (**Scheffler, et al., 2023**), funding agencies and industry partners are pushing for broader accessibility, reusability, and dissemination of APT data across academia, industry, and the public (**European Commission, 2018; Brainard, 2022; Fair data guide, 2024; White House, 2022**). This shift promises benefits such as richer contextualization, better understanding, minimized knowledge loss, secure data sharing, and greater automation in data analysis—helping address both quantitative and qualitative uncertainties in APT. However, the APT community still struggles with how to collectively embrace and address these challenges.

Most standardisation efforts so far have focused on the acquisition phase (**Miller, et al., 2006; Hudson, et al., 2011; Blum, et al., 2017; Marquis, et al., 2016; Exertier, et al., 2018; Gault, et al., 2024**) with less attention given to data analysis and research data management (**Kühbach, et al., 2022b**). In parallel, ontologies—ranging from simple term lists to taxonomies with complex relationships (**Gruber, 1993; 1995; 2009**)—can clarify the meaning of specific metadata and results in greatly improving machine-based, automatable reasoning in APT. The International Field Emission Society (IFES) has led the creation of a glossary and is working on an ISO standard (ISO 18115-1:2023). The consensus on developing standards necessitated converging on globally accepted terminologies and definitions as a starting point. At present, the most important 14 terms and developed definitions are now available in ISO TC201 terminology 18155-1 documentation. In the near future, in addition to the development of globally accepted terminology, additional topics including the development of standardised approaches and procedures for sample preparation, for recording and reporting APT experimental conditions, for reconstruction and visualisation of APT data are being discussed. The development of unified data type to enhance the shareability of datasets across research groups, and the development of guidelines for reporting of APT data in publications are also being considered. In parallel, comprehensive semantic models like NeXus, proposed by a joint effort of the FAIRmat consortium (**Jemian, et al., 2025**) and the NeXus International Advisory Committee, are promising.

The FAIR data management principles (**Wilkinson, et al., 2016; Jacobsen, et al., 2020; Barker, et al., 2022; Wilkinson, et al., 2025**) offer best practices for designing infrastructures that enable data reuse, not only for datasets but also for software and workflows. The core aim of FAIR is to make data more accessible and processable by machines, demanding advanced semantic frameworks and software interfaces that surpass current capabilities in APT (**Atom probe software tools, 2024**).

File formats like POS, APT, and RRNG/RNG are widely used and provide basic standardisation, but lack essential context on data provenance and processing. The redundant development of format readers could be avoided with shared libraries (**Kühbach, 2024b**), particularly when combined with libatomprobe and cross-language support. NeXus-based data models and their implementations (**Kühbach, et al., 2022b; Jemian, et al., 2025; Kühbach, 2024d**) present a more robust, open-source, and fully documented solution. Similarly, many APT groups use diverse electronic lab notebooks and research data management systems (**Barillari, 2016; Electronic lab notebooks, 2024; eLabFTW, 2024; The University of Sydney, 2021; Brandt, et al., 2021; Scheidgen, et al., 2023**), but interoperability remains an issue. Routine APT data analysis is now possible on local and high-performance computing systems, but past global database efforts faced challenges with file format support and user engagement (**Kühbach, 2022b; Ceguerra, et al., 2014; Diercks, et al., 2017; Meier, et al., 2023; Heller, et al., 2024**). New frameworks, such as NOMAD, address many technical barriers, yet widespread adoption depends on building consensus within the community, particularly due to the broad diversity of materials analysed. Additionally, legal and data availability constraints may restrict access to some datasets.

The main challenge lies in encouraging APT researchers to test, adopt, and improve new solutions, and in motivating software development toward interoperable tools that integrate with industry platforms. To advance documentation and data sharing—essentially, to achieve standardisation—progress is needed both in community consensus-building and technical development. Future efforts should focus on open, inclusive decision-making—ideally facilitated by organisations such as IFES.

## 4. Summary and outlook

The field of APT analysis has continued to progress significantly, with vibrant research activities in a number of areas. Crucial community engagement also remains required to continue developing APT as a scientific characterisation technique.

There remains much scope for detailed research on the mechanism of FEV, particularly for research using modern quantum-mechanical tools —notably density functional theory (DFT) but possibly also the Hellman-Feynman approach to evaluating electrostatic-field induced forces, e.g., (**Feynman, 1939**). Specific research aspects that may have immediate impacts include: transparent validation of DFT methods, as used in high-electrostatic-field situations; use of DFT theory to establish the configuration of the "geometrically ideal escape path" for an escaping ion nucleus and to follow the evolution of the partial charge on the atom along this path; re-investigation into how to define and calibrate the various slightly different physical electrostatic (ES) fields that are relevant to real emitter surfaces and their modelling. Corrections of several known theoretical errors in the Kingham theory of field post-ionisation (FPI) are needed. An investigation of the apparent incompatibilities (identified above) between the BBF mechanism and experiments is also needed, particularly the significant discrepancies between calculated "critical evaporation fields" and measured "operating fields". Specifically, could all historical APT field calibrations be seriously incorrect? We note here that the APT community uses a reference table, containing ZBEF values for pure elements, that is incomplete and inaccurate. It is incomplete because the field sensitivity values are needed. Also, the quoted ZBEF values are inaccurate because some known contributory effects have not been taken into account (**Forbes, 1995; Miller and Forbes, 2014**). Now, using DFT models, more accurate values can be estimated (**Ashton, et al., 2020; Ohnuma, 2021; Qi, et al., 2023**). This also can be done empirically, using field estimations, made possible with the systematic use of in-situ TEM-APT experiments, and the correlation with the Kingham estimation of local electric field (**Da Costa, et al., 2024; Kingham, 1982; Tegg, 2024**) (or with a future corrected estimation of these fields).

In terms of data analysis and reconstruction and analysis, our understanding of artifacts in reconstructed data continuously evolve. Modelling remains particularly challenging, both in terms of underlying physics but also the required computation power to run models. Simplified models, which are computationally more efficient and widely integrated into reconstruction software, may be a promising first step.

The APT community is innovative and thrives in developing new data analysis methods. Continued caution against the use of tools as a 'black box' is encouraged. We also suggest the creation of a centralised repository to share tools made in different labs. The development of AI/ML tools could also be both revolutionary (in terms of less subjectively determining analysis parameters) and dangerous (a true 'black box' and only as good as it has been trained) in equal measure, hence, the need for standards.

Progress towards widely adopted standards and best practices remains a major challenge for the APT community. However, standards are necessary to promote confidence in APT measurements and to make recommended practices known and accessible. The development of reference materials, consensus benchmarking values and tools, and reference data sets can form a foundation upon which accuracy, repeatability (i.e., between measurements), and reproducibility (i.e., between tools and labs) can be evaluated and improved. e.g., through calibration. Standards also provide the means to ground truth methods and models and to determine detection limits. The first steps may simply be about defining guidelines for data acquisition and full reporting of acquisition conditions and methods and parameters used for data analysis. Other suggestions include regular training sessions for new APT users in community-agreed best practices. This is an advantage of a research community that has substantial fewer members than for example, the electron microscopy or computational materials modelling.

**Disclaimers**

- Certain equipment, instruments, software, or materials are identified in this paper in order to specify the experimental procedure adequately. Such identification is not intended to imply recommendation or endorsement of any product or service by NIST, nor is it intended to imply that the materials or equipment identified are necessarily the best available for the purpose.

- These opinions, recommendations, findings, and conclusions do not necessarily reflect the views or policies of NIST or the United States Government.

**Acknowledgements**

Financial support of AFOSR – Ali Sayir for the organization of the workshop. August 2024.

EM acknowledges support from the U.S. Air Force Office of Scientific Research, award number FA9550-24-1-0091.

FV acknowledges support of the French Agence Nationale de la Recherche (ANR), under grant ANR-21-CE42-0024 (project HiKEAP).

AD acknowledges support from the Department of Energy (DOE), Office of Science, Basic Energy Sciences, Materials Sciences, and Engineering Division as part of the Early Career Research Program FWP 76052.

BM acknowledges support from National Science Foundation CAREER award number 2145091.

DKS acknowledges support from the U.S. Department of Energy (DOE), Office of Science, Basic Energy Sciences, Materials Sciences and Engineering Division, Mechanical Behavior, and Radiation Effects (FWP 56909). PNNL is a multiprogram national laboratory operated by Battelle for the U.S. DOE under Contract DE-AC05-79RL01830.

MK acknowledges support from the Deutsche Forschungsgemeinschaft (DFG, German Research Foundation) – 460197019 (FAIRmat).

## Bibliography

Ashton, M., Mishra, A., Neugebauer, J. & Freysoldt C. (2020). Ab Initio Description of Bond Breaking in Large Electric Fields. Physical Review Letters 124(17), 176801. https://doi.org/10.1103/PhysRevLett.124.176801.

Barker, M., Chue Hong, N. P., Katz, D. S., Lamprecht, A.-L., Martinez-Ortiz, C., Psomopoulos, F., Harrow, J., Castro, L.J., Gruenpeter, M., Martinez, P.A. & Honeyman T. (2022). Introducing the fair principles for research software. *Scientific Data*, 9(1), 622/ https://doi.org/10.1038/s41597-022-01710-x.

Bas, P., Bostel, A. Deconihout, B. & Blavette. D. (1995). A General Protocol for the Reconstruction of 3D Atom Probe Data. *Applied Surface Science* 87–88, 298–304. https://doi.org/10.1016/0169-4332(94)00561-3.

Bhuiyan, A.F.M., Feng, X., Johnson, J.M., Huang, H.-L., Sarker, J., Zhu, M., Karim, M.R., Mazumder, B., Hwang, J. & Zhao, H. (2020). Phase transformation in MOCVD growth of (AlxGa1-x)2O3 thin films, APL Materials 8, 031104. https://doi.org/10.1063/1.5140345

Biswas, R. K. & Forbes, R. G. (1982). Theoretical arguments against the Müller-Schottky mechanism of field evaporation. J. Phys. D: Appl. Phys. 15, 1323–1338.

https://doi.org/10.1088/0022-3727/15/7/026 Blum, T. B., Darling, J. R. Kelly, T. F., Larson, D. J., Moser, D. E., Perez-Huerta, A., Prosa, T. J., Reddy, S. M., Reinhard, D. A., Saxey, D. W., Ulfig, R. M. & Valley, J. W. (2017). Best Practices for Reporting Atom Probe Analysis of Geological Materials, in *Microstructural Geochronology: Planetary Records Down to Atom Scale*, D. E. Moser, F. Corfu, J. R. Darling, S. M. Reddy and K. Tait, Eds., Wiley, pp. 369-373. https://doi.org/10.1002/9781119227250.ch18.

Boyes, E. D., Waugh, A. R., Turner, P. J., Mills, P. F. & Southon, M. J. (1977) Abstracts of the 24th Int. Field Emission Symposium, Oxford (unpublished).

Brandon, D. G. (1966). On Field Evaporation. *Philosophical Magazine.* 14(130), 803–20. https://doi.org/10.1080/14786436608211973.

Brandt, N., Griem, L., Herrmann, C., Schoof, E., Tosato, G., Zhao, Y., Zschumme, P. & Selzer, M. (2021). Kadi4mat: A research data infrastructure for materials science. *Data Science Journal*, 20, 8. DOI: 10.5334/dsj-2021-008

Caballero, F., Miller, M. K., Babu, S. S. & Garcia-Mateo, C. (2007). Atomic scale observations of bainite transformation in a high carbon high silicon steel, Acta Mater 55(1), 381-390. doi: https://doi-org/10.1016/j.actamat.2006.08.033 .

Cadel, E., Vurpillot, F., Larde, R., Duguay, S. & Deconihout, B. (2009). Depth resolution function of the laser assisted tomographic atom probe in the investigation of semiconductors, J Appl Phys 106 (4), 044908. https://doi-org/10.1063/1.3186617 .

Car, R. & Parrinello, M. (1985). Unified Approach for Molecular Dynamics and Density-Functional Theory, Phys. Rev. Lett., 55(22), 2471–2474. https://doi-org/10.1103/PhysRevLett.55.2471.

Castin, N., Klups, P., Konstantinovic, M. J., Bonny, G., Pascuet, M. I., Moody, M. & Malerba, L. (2025). How precisely are solute clusters in RPV steels characterized by atom probe experiments?, *Journal of Nuclear Materials* 603, 155412. https://doi-org/10.1016/j.jnucmat.2024.155412

Ceguerra, A., Liddicoat, P., Apperley, M., Ringer, S. & Goscinski, W. (2014). Atom probe workbench version 1.0.0: An australian cloud based platform for the computational analysis of data from an atom probe microscope (apm), used for chemical and 3d structural materials charac terisation at the atomic scale. the workbench has been developed as part of the nectar-funded characterisation virtual laboratory.

Cerezo, A., Godfrey, T.J., Sijbrandij, S.J. Smith, G.D.W. & Warren, P.J. (1998) Performance of an energy-compensated three dimensional atom probe, Rev. Scientific Instruments 69, 49-58. https://doi.org/10.1063/1.1148477

Da Costa, G., Castro, C., Normand, A., Vaudolon, C., Zakirov, A.,Macchi, J., Ilham, M., Edalati, K., Vurpillot, F. & Lefebvre, W. (2024) Bringing atom probe tomography to transmission electron microscopes. Nat Commun 15, 9870. https://doi.org/10.1038/s41467-024-54169-2.

De Geuser, F., Lefebvre, W., Danoix, F., Vurpillot, F., Forbord, B. and Blavette, D. (2007), An improved reconstruction procedure for the correction of local magnification effects in three-dimensional atom-probe. Surf. Interface Anal., 39: 268-272. https://doi.org/10.1002/sia.2489

De Geuser, F., Gault, B. Bostel. A. & Vurpillot, F. (2007). Correlated Field Evaporation as Seen by Atom Probe Tomography. *Surface Science* 601(2), 536–43. https://doi.org/10.1016/j.susc.2006.10.019.

DeRocher, K., McLean, M., Meisenkothen, F. (2022) A Standards-based Approach to Dopant Quantification Using Atom Probe Tomography, Microscopy and Microanalysis, 28, S1, 728–729, https://doi.org/10.1017/S1431927622003373

Diercks, D.R., Gorman, B.P., & Gerstl, S.A. (2017). An open access atom probe tomography mass spectrum database. *Microscopy and Microanalysis*, 23(S1):664–665. https://doi.org/10.1017/S1431927617003981

Di Russo, E., F. Moyon, N. Gogneau, L. Largeau, E. Giraud, J.-F. Carlin, N. Grandjean, Chauveau J. M., Hugues M., Blum I., Lefebvre W., Vurpillot F., Blavette D. & Rigutti L. (2018). Composition Metrology of Ternary Semiconductor Alloys Analyzed by Atom Probe Tomography. *The Journal of Physical Chemistry C* 122 (29): 16704–14. https://doi.org/10.1021/acs.jpcc.8b03223.

Dong, Y., Etienne, A., Frolov, A., Fedotova, S., Fujii, K., Fukuya, K., Hatzoglou, C., Kuleshova, E., Lindgren, K., London, A., Lopez, A., Lozano-Perez, S.., Miyahara, Y., Nagai, Y., Nishida, K., Radiguet, B., Schreiber, D., Soneda, N., Thuvander, M., Toyama, T., Sefta, F., Wang, J., Chou, P. & Marquis, E.A. (2019) Atom Probe Tomography Interlaboratory Study on Clustering Analysis in Experimental Data Using the Maximum Separation Distance Approach, *Microscopy and Microanalysis,* vol. 25, no. 2, pp. 356-366. https://doi.org/10.1017/S1431927618015581

Eyring, H. (1935) The Activated Complex in Chemical Reactions, *The Journal of Chemical Physics*, vol. 3, no. 2, pp. 107–115, https://doi.org/10.1063/1.1749604.

Electronic lab notebooks, 2024.

eLabFTW, 2024. https://www.elabftw.net/

European Commission: Directorate-General for Research and Innovation & PwC EU Services. (2018). Cost-benefit analysis for FAIR research data: cost of not having FAIR research data. Publications Office. https://data.europa.eu/doi/10.2777/02999.

Exertier, F., La Fontaine, A., Corcoran, C., Piazolo, S., Belousova, E., Peng, Z., Gault, B., Saxey, D.W., Fougerouse, D., Reddy, S.M., Pedrazzini, S., Bagot, P.A.J., Moody, M.P., Langelier, B., Moser, D.E., Botton, G.A., Vogel, F., Thompson, G.B., Blanchard, P.T.,

Chiaramonti, A.N., Reinhard, D.A., Rice, K.P., Schreiber, D.K., Kruska, K., Wang, J. & Cairney, J.M. (2018) Atom probe tomography analysis of the reference zircon gj-1: An interlaboratory study. *Chemical Geology*, 495:27–35.

https://doi-org.proxy.lib.umich.edu/10.1016/j.chemgeo.2018.07.031

Fair data guide, 2024.

Feynman, R. P. (1939) Forces in Molecules. Phys. Rev. 56, 340–343.
https://doi.org/10.1103/PhysRev.56.340

Forbes, R. G. (1995) Field Evaporation Theory: A Review of Basic Ideas. *Applied Surface Science*, 87/88:1–11. https://doi.org/10.1016/0169-4332(94)00526-5.

Forbes, R. G. (2024). On the scientific status of the theory of field desorption and field evaporation. 2024 Workshop on Frontiers of APT Physics, Data Processing, Analysis and Reconstruction, Arlington, Virginia, USA, August 2024.
https://doi.org/10.13140/RG.2.2.25625.53602

Galtrey, M.J., Oliver, R.A., Kappers, M.J., Humphreys, C.J., Stokes, D.J., Clifton, P.H. & Cerezo, A. (2007) Three-dimensional atom probe studies of an InxGa1-xN/GaN multiple quantum well structure: Assessment of possible indium clustering, Appl Phys Lett, 90, 6.
https://doi.org/10.1063/1.2431573.

Gault, B., de Geuser, F., Stephenson, L., Moody, M.P., Barrington, B.C., Muddle, S. & Ringer, S.P. (2008). Estimation of the Reconstruction Parameters for Atom Probe Tomography. *Microscopy and Microanalysis* 14 (4): 296–305.
https://doi.org/10.1017/S1431927608080690.

Gault, B., Marquis, E.A., Saxey, D.W., Hughes, G.M., Mangelinck, D., Toberer, E.S. & Snyder, G.J. (2010) High-resolution nanostructural investigation of Zn4Sb3 alloys, Scripta Mater, 63(7) 784-787. https://doi.org/10.1016/j.scriptamat.2010.06.014

Gault, B., Chiaramonti, A., Cojocaru-Mirédin, O., Stender, P., Dubosq, R., Freysoldt, C., Makineni, S.K., Li, T., Moody, M. & Cairney, J.M. (2021). Atom Probe Tomography. *Nature Reviews, Methods Primers*, 1–51. https://doi.org/10.1038/s43586-021-00047-w

Gault, B., Saksena, A., Sauvage, X., Bagot, P., Aota, L.S., Arlt, J., Belkacemi, L.T., Boll, T., Chen, Y.-S., Daly, L., Djukic, M.B., Douglas, J.O., Duarte, M.J., Felfer, P.J., Forbes, R.G., Jing Fu, J., Gardner, H.M., Gemma, R., Gerstl, S.A., Gong, Y., Hachet, G., Jakob, S., Jenkins, B.M., Jones, M.E., Khanchandani, H., Kontis, P., Kr¨amer, M., Kühbach, M.,

Marceau, R.K.W., Mayweg, D., Moore, K.L., Nallathambi, V., Ott, B.C., Poplawsky, J.D., Prosa, T., Pundt, A., Saha, M., Schwarz, T.M., Shang, Y., Shen, X., Vrellou, M., Yu, Y., Zhao, Y., Zhao, H. and Zou, B. (2024) Towards establishing best practice in the analysis of hydrogen and deuterium by atom probe tomography. Microscopy & Microanalysis 30(6) 1205-1220. https://doi.org/10.1093/mam/ozae081

Ghamarian, I., Marquis, E.A. (2019) Hierarchical Density-Based Cluster Analysis Framework for Atom Probe Tomography Data, Ultramicroscopy. 200 28-38 https://doi.org/10.1016/j.ultramic.2019.01.011

Ghamarian, I., Marquis, E.A. (2020) Morphological classification of dense objects in atom probe tomography data, Ultramicroscopy 215 112996 https://doi.org/10.1016/j.ultramic.2020.112996

Ghamarian, I., Yu, L.-J., Marquis, E.A. (2020b) Quantification of solute topology in atom probe tomography data: application to the microstructure of a proton-irradiated Alloy 625, Metall. Mater. Trans 51 42-50 https://doi.org/10.1007/s11661-019-05520-6

Gomer, R. (1959) Field Desorption. *The Journal of Chemical Physics* 31 (2): 341–45. https://doi.org/10.1063/1.1730354.

Gopon, P., Douglas, J.O., Meisenkothen, F., Singh, J., London, A.J., Moody, M.P. (2022) Atom Probe Tomography for Isotopic Analysis: Development of the 34S/32S System in Sulfides, Microscopy and Microanalysis, 28 (4) 1127–1140, https://doi.org/10.1017/S1431927621013568

Gruber. T. (2009) *Ontology*, pages 1963–1965. Springer US. https://doi-org/10.1007/978-0-387-39940-9_1318

Gruber, T.R. (1993) A translation approach to portable ontology specifications. *Knowledge Acquisition*, 5(2):199–220. https://doi-org/10.1006/knac.1993.1008

Gruber, T.R. (1995). Toward principles for the design of ontologies used for knowledge sharing. *International Journal of Human-Computer Studies*, 43(5–6):907–928. https://doi-org/10.1006/ijhc.1995.1081

Haley, D.J., Choi, P. & Raabe, D. (2015) Guided mass spectrum labelling in atom probe tomography, *Ultramicroscopy*, vol. 159, pp. 338-345. https://doi-org/10.1016/j.ultramic.2015.03.005

Hatzoglou, C., Radiguet, B. & Pareige, P. (2017). Experimental Artefacts Occurring during Atom Probe Tomography Analysis of Oxide Nanoparticles in Metallic Matrix: Quantification and Correction. *Journal of Nuclear Materials* 492, 279–91.

https://doi.org/10.1016/j.jnucmat.2017.05.008.

Hatzoglou, C, Da Costa, G. & Vurpillot, F. (2019). Enhanced Dynamic Reconstruction for Atom Probe Tomography. *Ultramicroscopy* 197, 72–82.

https://doi.org/10.1016/j.ultramic.2018.11.010 .

Hatzoglou, C., Rouland, S., Radiguet, B., Etienne, A., Da Costa, G., Sauvage, X., Pareige, P. & Vurpillot, F. (2020). Preferential Evaporation in Atom Probe Tomography: An Analytical Approach. *Microscopy and Microanalysis* 26 (4): 689–98. https://doi.org/10.1017/S1431927620001749.

Heller, M., Ott, B., Dalbauer,V. & Felfer, P. (2024) A mat lab toolbox for findable, accessible, interoperable, and reusable atom probe data science. *Microscopy and Microanalysis*, 30(6) 1138-1151. https://doi.org/10.1093/mam/ozae031

Hudson, D., Smith, G.D.W. & Gault, B. (2011) Optimisation of mass ranging for atom probe microanalysis and application to the corrosion processes in Zr alloys," Ultramicroscopy, vol. 111, no. 6, pp. 480-486. https://doi.org/10.1016/j.ultramic.2010.11.007 .

Hyde, J.M. & English, C. A. (2001) An Analysis of the Structure of Irradiation Induced Cu-Enriched Clusters in Low and High Nickel Welds, in *Materials Research Society Symposium Proceedings.* https://doi-org/10.1557/PROC-650-R6.6

Hyde, J.M., Burke, M.G., Gault, B., Saxey, D.W., Styman, P.D., Wilford, K.W. & Williams, T.J. (2011) Atom probe tomography of reactor pressure vessel steels: An analysis of data integrity, *Ultramicroscopy,* 111, 676-682. https://doi-org/10.1016/j.ultramic.2010.12.033

Hyde, J.M., Costa, G.D., Hatzoglou, C., Weekes, H., Radiguet, B., Styman, P.D., Vurpillot, F., Pareige, C., Etienne, A., Bonny, G., Castin, N., Malerba, L. & Pareige, P. (2017) Analysis of Radiation Damage in Light Water Reactors: Comparison of Cluster Analysis Methods for the Analysis of Atom Probe Data, *Microscopy and Microanalysis,* 23(2) 366-375. https://doi-org/10.1017/s1431927616012678

Iso 18115-1:2023(en) surface chemical analysis — vocabulary — part 1: General terms and terms used in spectroscopy. 2023

IUPAC (2019). Gold Book, Compendium of chemical terminology. Version 5.0.0. DOI: 10.1351/goldbook

Jacobsen, A., de Miranda Azevedo, R., Juty, N., Batista, D., Coles, S., Cornet, R., Courtot, M., Crosas, M., Dumontier, M., T. Evelo, C.T., Goble, C., Guizzardi, G., Hansen, K.K., Hasnain, A., Hettne, K., Heringa, J., Hooft, R.W.W., Imming, M., Jeffery, K.G., Kaliyape rumal, R., Kersloot, M.G., Kirkpatrick, C.R., Kuhn, T., Labastida, I., Magagna, B., McQuilton, P., Meyers, N., Montesanti, A., van Reisen, M., Rocca-Serra, P., Pergl, R., Sansone, S.-A., da Silva Santos, L.O.B., Schneider, J., Strawn, G., Thompson, M., Waagmeester, A., Weigel, T., Wilkinson, M.D., Willighagen, E.L., Wittenburg, P., Roos, M., Mons, B. & Schultes. E. (2020) Fair principles: Interpretations and implementation considerations. *Data Intelligence*, 2(1–2):10–29

Jemian, P.R. Pielsticker, L., Akeroyd, F., Richter, T., De Nolf, W., Chang, P., Brewster, A.S., Dobener, F., sanbrock, Wuttke, J., Bernstein, H.J., Peterson, P., Watts, B., mkoennecke, RussBerg, Kühbach, M., Wintersberger, E., Osborn, R., atomprobe-tc, cmmngr, RubelMozumder, Shabih, S., Campbell, S., Kienzle, P., Jones, D., Caswell, T.A., Albino, A., Clarke, M., soph-dec, Rettig, L. (2025). nexusformat/definitions: v2025.11 (v2025.11). Zenodo. https://doi.org/10.5281/zenodo.17607931

Jenkins, B. M., Zakirov, A., Vurpillot, F., Etienne, A., Pareige, C., Pareige, P., & Radiguet, B. (2024) On the Iron Content of Mn-Ni-Si-Rich Clusters That Form in Reactor Pressure Vessel Steels during Exposure to Neutron Irradiation. *Acta Materialia* 281, 120384. https://doi.org/10.1016/j.actamat.2024.120384

Khan, M.A., Ringer, S.P. & Zheng, R.K. (2016) Atom Probe Tomography on Semiconductor Devices, Adv Mater Interfaces, 3(12), 1500713 https://doi.org/10.1002/admi.201500713

Kelly, T.F. & D. J. Larson, D.J. (2012) Atom Probe Tomography, Annu Rev Mater Res, vol. 42, 1-31 doi: https://doi.org/10.1146/annurev-matsci-070511-155007

Kingham, D R. (1982) The Post-Ionization of Field Evaporated Ions: A Theoretical Explanation of Multiple Charge States. *Surface Science* 116 (2): 273–301. https://doi.org/10.1016/0039-6028(82)90434-4

Kinno, T., Akutsu, H., Tomita, M., kawanaka, S., Sonehara, T., Hokazono, A., Renaud, L., Martin, I., benbalagh, R., Salle, B., Takeno, S. (2012) Influence of multi-hit capability on quantitative measurement of NiPtSi thin film with laser-assisted atom probe tomography, Appl Surf Sci, 259, 726-730, doi: https://doi.org/10.1016/j.apsusc.2012.07.108

Kühbach, M., Bajaj, P., Zhao, H., Murat H. C¸ A. Jagle, E.A., and Gault, B. (2021) On strong-scaling and open-source tools for analyzing atom probe tomography data. *npj Computational Materials*, 7(1). https://doi.org/

Kühbach, M., Rielli, V.V., Primig, S., Saxena, A., Mayweg, A., Jenkins, B., Antonov, S., Reichmann, A., Kardos, S., Romaner, L., and Brockhauser, S. (2022a) On strong-scaling and open-source tools for high-throughput quantification of material point cloud data: Composition gradients, microstructural object reconstruction, and spatial correlations https://doi.org/10.48550/arXiv.2205.13510

Kühbach, M. London, A.J., Wang, J., Schreiber, D.K., Martin, F.M., Ghamarian, I., Bilal, H., and Ceguerra, A.V. (2022b) Community-driven methods for open and reproducible software tools for analyzing datasets from atom probe microscopy. *Microscopy and Micro analysis*, 28(4):1038–1053. https://doi.org/10.1017/S1431927621012241

Kühbach, M. Menon, S., Saxena, A., Forti, M., Kruˇzˊıkovˊa, P., Hammerschmidt, T., Draxl, C., and Hickel, T., (2024a). Nfdi-matwerk iuc 09: Infrastructure interfaces with condensed matter physics (collaboration with fairmat). https://doi.org/10.5281/zenodo.12594062

Kühbach, M. (2024b) https://atomprobe-tc.github.io/ifes_apt_tc_data_modeling/

Kühbach, M. S. Brockhauser, F. Dobener, R. Mozumder, L. Pielsticker, S. Shabih, H. Weber, C. Koch, and C. Draxl. (2024c) https://fairmat-nfdi.github.io/pynxtools-apm/

Larson, D. J., Prosa, T.J., Ulfig, R.M., Geiser, B.P., and Kelly, T.F. (2013) *Local Electrode Atom Probe Tomography*. New York, NY: Springer New York. https://doi.org/10.1007/978-1-4614-8721-0.

Li, Y., Zhou, X., Colnaghi, T., Colnaghi, T., Wei, Y., Marek, A., Li, H., Bauer, S., Rampp, M. & Stephenson, L.T.. Convolutional neural network-assisted recognition of nanoscale L12 ordered structures in face-centred cubic alloys. npj Comput Mater 7, 8 (2021). https://doi-org/10.1038/s41524-020-00472-7

Li, Y., Ye Wei, Y., Alaukik Saxena, A., Markus Kühbach, M., Freysoldt, C., Gault, B. (2026) Machine learning enhanced atom probe tomography analysis. Progress in materials Science., 156, 101561. https://doi-org/10.1016/j.pmatsci.2025.101561

Licata, O.G., Broderick, S., Rocco, E., Shahedipour-Sandvik, F., Mazumder, B.(2021) Dopant-defect interactions in Mg doped GaN via atom probe tomography", Appl Phys Lett., 119, 032102. https://doi.org/10.1063/5.0061153

London, A.J., Haley, D.J. and Moody, M.P. (2017) Single-Ion Deconvolution of Mass Peak Overlaps for Atom Probe Microscopy, *Microscopy and Microanalysis,* vol. 23, no. 2, pp. 300-306. https://doi.org/10.1017/S1431927616012782

Mancini, L., Amirifar, N., Shinde, D., Blum, I., Gilbert, M., Vella, A., Vurpillot, F., Lefebvre, W., Larde, R., talbot, E., Pareige, P., Portier, X., Zinai, A., Davesnne, C., Durand, C., Eymeri, J., Butte, R., Carlin, J.-F., Grandjean, N., Rigutti, L. (2014) Composition of Wide Bandgap Semiconductor Materials and Nanostructures Measured by Atom Probe Tomography and Its Dependence on the Surface Electric Field. *The Journal of Physical Chemistry C* 118 (41): 24136–51. https://doi.org/10.1021/jp5071264

Marquis, E. A., Geiser, B. P. Prosa, T. J. & Larson, D. J. (2011). Evolution of Tip Shape during Field Evaporation of Complex Multilayer Structures. *Journal of Microscopy* 241 (3): 225–33. https://doi.org/10.1111/j.1365-2818.2010.03421.x .

Marquis, E A., & Vurpillot, F. (2008). Chromatic Aberrations in the Field Evaporation Behavior of Small Precipitates. *Microscopy and Microanalysis* 14 (6): 561–70. https://doi.org/10.1017/S1431927608080793.

Marquis, E.A., Araullo-Peters, V., Etienne, A., Fedotova, S., Fujii, K., Fukuya, K., Kuleshova, E., Legrand, A., London, A., Lozano-Perez, S., Nagai, Y., Nishida, K., Radiguet, B., Schreiber, D., Soneda, N., Thuvander, M., Toyama, T., Sefta, F. & Chou, P. (2016) A round robin experiment: Analysis of solute clustering from atom probe tomography data. *Microscopy and Microanalysis*, 22(S3):666–667.

Marquis, E.A., Araullo-Peters, V., Dong, Y., Etienne, A., Fedotova, S., Fujii, K., Fukuya, K., Kuleshova, E., Lopez, A. London, A., Lozano-Perez, S., Nagai, Y., Nishida, K., Radiguet, B., Schreiber, D., Soneda, N., Thuvander, M., Toyama, T., Sefta, F. & Chou, P. (2018) On the Use of Density-Based Algorithms for the Analysis of Solute Clustering in Atom Probe Tomography Data, in *Proceedings of the 18th International Conference on Environmental Degradation of Materials in Nuclear Power Systems – Water Reactors*, 2018

McKinstry, D. 1972. 'An Examination of Field Evaporation Theory'. *Surface Science* 29 (1): 37–59. https://doi.org/10.1016/0039-6028(72)90070-2.

Meier, M.S., Bagot, P.A.J., Moody, M.P. and Haley, D. (2023) Large-scale atom probe tomography data mining: Methods and application to inform hydrogen behavior. *Microscopy and Microanalysis*, 29(3):879– 889 https://doi.org/10.1093/micmic/ozad027

Meisenkothen, F., Steel, E.B., Prosa, T.J., Henry, K.T., and Kolli. R.P. (2015) Effects of Detector Dead-Time on Quantitative Analyses Involving Boron and Multi-Hit Detection Events in Atom Probe Tomography'. *Ultramicroscopy* 159, 101–11. https://doi.org/10.1016/j.ultramic.2015.07.009.

Meisenkothen, F., Samarov, D.V., Kalish, I, Steel, E.B. (2020) Exploring the accuracy of isotopic analyses in atom probe mass spectrometry, Ultramicroscopy 216, 113018 https://doi.org/10.1016/j.ultramic.2020.113018

Meisenkothen, F., McLean, M., Kalish, I., Samarov, D.V., Steel, E.B. (2020b) Atom probe mass spectrometry of uranium isotopic reference materials, Anal. Chem. 2020, 92, 16, 11388–11395. https://doi-org/10.1021/acs.analchem.0c02273

Menand, A. & Blavette, D. (1986). Temperature dependence of iridium field evaporation rate. J. de Phys. Colloque 47 (C7), c7-17–C7-20.  https://doi.org/10.1051/jphyscol:1986704

Mikhalychev, A., Svlasenko, S., Payne, T. R., Reinhard, D. A., and Ulyanenkov, A. (2020) Bayesian approach to automatic mass-spectrum peak identification in atom probe tomography,"*Ultramicroscopy*, vol. 215, p. 113014 https://doi-org/10.1016/j.ultramic.2020.113014

Miller, M.K., Liu, C.T., Wright, J.A., Tang, W. and Hildal. K. (2006) APT characterization of some iron-based bulk metallic glasses. *Intermetallics*, 14(8):1019– 1026, 2006. Fourth International Conference on Bulk Metallic Glasses https://doi-org/10.1016/j.intermet.2006.01.040

Miller, M.K. & Forbes, R.G. "Field Evaporation and Related Topics," *Atom-Probe Tomography: The Local Electrode Atom Probe*. Springer US, Boston, MA, pp. 111–187, 2014. https://doi-org/10.1007/978-1-4899-7430-3_3

Morris, R.J.H., Cuduvally, R., Melkonyan, D., Fleischmann, C., Zhao, M., Arnoldi, L., van der Heide, P., Vandervorst, W. (2018) Toward accurate composition analysis of GaN and AlGaN using atom probe tomography, J Vac Sci Technol B, 36(3), 03f130 https://doi-org/10.1116/1.5019693.

Muller, M., Saxey, D.W., Smith, G.D.W. and Gault, B. (2011) Some aspects of the field evaporation behaviour of GaSb, Ultramicroscopy, 111(6) 487- 492,  https://doi-org/10.1016/j.ultramic.2010.11.019

Müller, E W. (1941) 'Abreißen adsorbierter Ionen durch hohe elektrische Feldstßrken'. *Die Naturwissenschaften* 29 (35): 533–34. https://doi.org/10.1007/BF01481175.

Müller, E W. (1956) 'Field Desorption'. *Physical Review* 102 (3): 618–24. https://doi.org/10.1103/PhysRev.102.618.

Ndiaye, S., Bacchi, C., Klaes, B., Canino, M., Vurpillot, F., and Rigutti. L. (2023) Surface Dynamics of Field Evaporation in Silicon Carbide. *The Journal of Physical Chemistry C* 127 (11): 5467–78. https://doi-org.proxy.lib.umich.edu/10.1021/acs.jpcc.2c08908

NIST Statistical Software (2025) https://www.nist.gov/itl/sed/products-services/statistical-software

Oberdorfer, C., Eich, S.M. & Schmitz, G. (2013). A Full-Scale Simulation Approach for Atom Probe Tomography. *Ultramicroscopy* 128, 55–67. https://doi.org/10.1016/j.ultramic.2013.01.005.

Oberdorfer, C., Withrow, T., Yu, L.-J., Fisher, K., Marquis, E.A. & Windl, W. (2018) Influence of surface relaxation on solute atoms positioning withing atom probe tomography reconstructions. Materials Characterization 146, 324-335. https://doi.org/10.1016/j.matchar.2018.05.014

Ohnuma, T. (2021) First-Principles Calculation of the Evaporation Field and Roll-up Effect of M (M = Fe, Cu, Si, and Mn) on the Fe (001) and Fe Step Structure. *Microscopy and Microanalysis*, March, 1–7. https://doi.org/10.1017/S1431927621000155 .

Barillari C., Ottoz D.S., Fuentes-Serna J.M., Ramakrishnan C., Rinn B., Rudolf F. (2016) openBIS ELN-LIMS: an open-source database for academic laboratories. Bioinformatics. 15;32(4):638-40 https://doi.org/10.1093/bioinformatics/btv606

Peng, Z., Vurpillot, F., Choi, P.-P., Li, Y., Raabe, D., and Gault, B. (2018) On the Detection of Multiple Events in Atom Probe Tomography. *Ultramicroscopy* 189 (June):54–60. https://doi.org/10.1016/j.ultramic.2018.03.018 .

Peng, Z., Zanuttini, D., Gervais, B., Jacquet, E., Blum, I., Choi, P.-P., Raabe, D., Vurpillot, F., Gault, B. (2019) Unraveling the Metastability of C-n(2+) (n=2-4) Clusters," J Phys Chem Lett, 10, 3, 581-588 https://doi.org/10.1021/acs.jpclett.8b03449 .

Pollak E. and Talkner, P. (2005) Reaction rate theory: What it was, where is it today, and where is it going?, *Chaos: An Interdisciplinary Journal of Nonlinear Science*, 15, 2, 026116 https://doi.org/10.1063/1.1858782 .

Qi, J. Oberdorfer, C. Windl, W. & Marquis, E.A. (2022) Ab initio simulation of field evaporation, *Phys. Rev. Materials*, vol. 6, no. 9, p. 093602. https://doi.org/10.1103/PhysRevMaterials.6.093602.

Qi, J., Oberdorfer, C., Marquis, E.A. & Windl, W. (2023a) Origin of enhanced zone lines in field evaporation maps, *Scripta Materialia*, vol. 230, p. 115406. https://doi.org/10.1016/j.scriptamat.2023.115406.

Qi, J., Xue, F., Marquis, E.A. & Windl, W. (2023b) Unexpected Field Evaporation Sequence in γ-TiAl: interpreting evaporation through dynamic bond breaking processes. Acta Materialia 287, 120741 https://doi-org/10.1016/j.actamat.2025.120741

Rigutti, L., Mancini, L., Hernández-Maldonado, D., Lefebvre, W., Giraud, E., Butté, R., Carlin, J.F., Grandjean, N., Blavette, D., and Vurpillot. F. (2016) Statistical Correction of Atom Probe Tomography Data of Semiconductor Alloys Combined with Optical Spectroscopy: The Case of Al0.25Ga0.75N. *Journal of Applied Physics* 119 (10): 105704. https://doi.org/10.1063/1.4943612 .

Ronsheim, P.A., Hatzistergos, M., and Jin, S. (2010) Dopant measurements in semiconductors with atom probe tomography," J Vac Sci Technol B, vol. 28, no. 1, pp. C1e1- C1e4 https://doi.org/10.1116/1.3242422 .

Ronsheim, P., Flaitz, P., Hatzistergos, M., Molella, C., Thompson, K., and Alvis, R. (2008) Impurity measurements in silicon with D-SIMS and atom probe tomography, Appl Surf Sci, 255, 4, 1547-1550 https://doi.org/10.1016/j.apsusc.2008.05.247

Rousseau, L., Normand, A., Morgado, F.F., Søreide, H.-S.M.S., Stephenson, L.T., Hatzoglou, C., Da Costa, G., Tehrani, K., Freysoldt, C., Gault, B. & Vurpillot, F. (2023) Introducing Field Evaporation Energy Loss Spectroscopy. *Communications Physics* 6 (1): 1–8. https://doi.org/10.1038/s42005-023-01203-2

Russo, E., Blum, I., Houard, J., Gilbert, M., Da Costa, G., Blavette, D., & Rigutti, L. (2018a) 'Compositional Accuracy of Atom Probe Tomography Measurements in GaN: Impact of Experimental Parameters and Multiple Evaporation Events'. *Ultramicroscopy* 187 126–34. https://doi.org/10.1016/j.ultramic.2018.02.001

Russo, E., Blum, I., Houard, J., Gilbert, M., Da Costa, G., Blavette, D., & Rigutti, L. (2018b) 'Compositional Accuracy of Atom Probe Tomography Measurements in GaN: Impact of Experimental Parameters and Multiple Evaporation Events'. *Ultramicroscopy* 187 126–34. https://doi.org/10.1016/j.ultramic.2018.02.001

Sánchez, C. G., Lozovoi, A. Y. & Alavi, A. (2004). Field-evaporation from first-principles. Mol Phys. 102, 1045–1055.  https://doi.org/10.1080/00268970410001727673

Sarker, J., Broderick, S., Bhuiyan, A.F.M.A.U., Feng, Z., Zhao, H., Mazumder, B.  (2020) A combined approach of atom probe tomography and unsupervised machine learning to understand phase transformation in (AlxGa1-x)2O3, Appl Phys Lett., 116, 152101. https://doi.org/10.1063/5.0002049

Saxena, A., Polin, N., Kusampudi, N., Katnagallu, S., Molina-Luna, L., Gutfleisch, O., Berkels, B., Gault, B., Neugebauer, J. & Freysoldt, C. (2023) A Machine Learning Framework for Quantifying Chemical Segregation and Microstructural Features in Atom Probe Tomography Data. Microscopy & Microanalysis, 29, 1658-1670. https://doi.org/10.1093/micmic/ozad086

Saxena, A., Kühbach, M., Katnagallu, S., Kontis, P., Gault, B.,  Freysoldt, C. Analyzing Linear Features in Atom Probe Tomography Datasets using Skeletonization, Microscopy and Microanalysis,  30, S1,  ozae044.022, https://doi.org/10.1093/mam/ozae044.022

Scheffler, M., Aeschlimann, M., Albrecht, M., Bereau, T., Bungartz, H.-J., Felser, C., Greiner, M., Groß, A., Koch, C.T., Kremer, K., Nagel, W.E., Scheidgen, M.,  Wöll, C., and Draxl. C. (2022) Fair data enabling new horizons for materials research. *Nature*, 604(7907):635–642. https://doi-org/10.1038/s41586-022-04501-x

Scheidgen, M., Himanen, L., Ladines, A.N., Sikter, D.,  Nakhaee, M., Fekete, A., Chang, T., Golparvar, A.,  M´arquez, J.,  Brockhauser, S., Br¨uckner, S., Ghiringhelli, L.M., Dietrich, F., Lehmberg, D., Denell, T., Albino, A., Näsström, H., Shabih, S., Dobener, F., Kühbach, M., Mozumder, R., Rudzinski, J.F., Daelman, N., Pizarro, J.M., Kuban, M., Salazar, C., Ondra˘cka, P., Bungartz, H.-J., and  Draxl. C (2023) Nomad: A distributed web-based platform for managing materials science research data. *Journal of Open Source Soft ware*, 8(90):5388. https://doi.org/10.21105/joss.05388

Silaeva, E.P., Houard, J., Hideur, A., Martel, G., and Vella, A. (2015) Field evaporation of semiconductors assisted by nonequilibrium phonon excitations, *Phys. Rev. B*, 92(19)  195307, https://doi-org/10.1103/PhysRevB.92.195307.

Smith, G.D.W. (1986) Field Ion Microscopy and Atom Probe Microanalysis, ASM Handbook, Metals Park, OH, 1986, pages 583-602. https://doi-org/10.31399/asm.hb.v10.a0001772

Tegg, L., Stephenson, L.T. and Cairney, J.M. (2024) Estimation of the Electric Field in Atom Probe Tomography Experiments Using Charge State Ratios. *Microscopy and Microanalysis*, 30(3) 466-475. https://doi.org/10.1093/mam/ozae047.

Tsong, T T. (1978) Field Ion Image Formation. *Surface Science* 70:211–33. https://doi-org/10.1016/0039-6028(78)90410-7

Tsong, T.T. (1978b) J. Phys. F: Met. Phys. 8 1349 https://doi.org/10.1088/0305-4608/8/7/008

Tsong, T.T., Mclane, S.B., and Kinkus, T.J. (1982) Pulsed-Laser Time-of-Flight Atom-Probe Field-Ion Microscope, Rev Sci Instrum, vol. 53, no. 9, pp. 1442-1448. https://doi.org/10.1063/1.1137193.

Tsong, T. T. (1990) Atom-Probe Field-Ion Microscopy (Cambridge Univ. Press, 1990).

The University of Sydney (2021) Atom probe data share. https://micro.org.au/online-tools/atom-probe-data-share/

Vineyard, G.H. (1957) Frequency factors and isotope effects in solid state rate processes, *Journal of Physics and Chemistry of Solids*, 3, 1, 121–127, https://doi.org/10.1016/0022-3697(57)90059-8.

Voter, A.F., Montalenti, F., and Germann, T.C. (2002) Extending the Time Scale in Atomistic Simulation of Materials, *Annual Review of Materials Research*, 32, 1, 321–346, https://doi.org/10.1146/annurev.matsci.32.112601.141541.

Vurpillot, F., Bostel, A., Menand, A., and Blavette, D. (1999) Trajectories of field emitted ions in 3D atom-probe, *The European Physical Journal - Applied Physics*, vol. 6, no. 2. pp. 217–221 https://doi.org/10.1051/epjap:1999173.

Vurpillot, F., Gault, B., Geiser, B.P., and Larson, D.J. (2013) Reconstructing Atom Probe Data: A Review. *Ultramicroscopy* 132:19–30. https://doi.org/10.1016/j.ultramic.2013.03.010.

Vurpillot, F., and Oberdorfer, C.. (2015) Modeling Atom Probe Tomography: A Review. *Ultramicroscopy* 159:202–16. https://doi.org/10.1016/j.ultramic.2014.12.013.

Vurpillot, F. (2016) Chapter Two - Field Ion Emission Mechanisms, in *Atom Probe Tomography*, W. Lefebvre-Ulrikson, F. Vurpillot, and X. Sauvage, Eds., Academic Press, 2016, pp. 17–72. https://doi.org/10.1016/B978-0-12-804647-0.00002-4.

Vurpillot, F., Bostel, A. and Blavette, D. (2000) Trajectory Overlaps and Local Magnification in Three-Dimensional Atom Probe. *Applied Physics Letters* 76 (21): 3127–29. https://doi.org/10.1063/1.126545.

Vurpillot, F., Hatzoglou, C., Klaes, B., Rousseau, L., Maillet, J.-B., Blum, I., Gault, B. and Cerezo, A. (2024) Crystallographic Dependence of Field Evaporation Energy Barrier in Metals Using Field Evaporation Energy Loss Spectroscopy Mapping. *Microscopy and Microanalysis*, 30:6, 1091-1099. https://doi.org/10.1093/mam/ozae083

Wada, M. (1984) On the Thermally Activated Field Evaporation of Surface Atoms. *Surface Science* 145:451–65. https://doi-org/10.1016/0039-6028(84)90093-1

Wang, A., Hatzoglou, C., Sneed, B., Fan, Z., Guo, W., Jin, K., Chen, D., et al. (2020) Interpreting Nanovoids in Atom Probe Tomography Data for Accurate Local Compositional Measurements. *Nature Communications* 11 (1): 1022. https://doi.org/10.1038/s41467-020-14832-w

Waugh, A.R., Boyes, E.D. and Southon, M.J. (1976) Investigations of Field Evaporation with a Field-Desorption Microscope. *Surface Science* 61 (1): 109–42. https://doi.org/10.1016/0039-6028(76)90411-8

Wei, J., Ulfig, R., Reinhard, D., Strennen, E., Larson, D.J., Voyles, P.M. (2025) Automated Mass Spectrum Labelling in Atom Probe Tomography, Microscopy and Microanalysis, 31 S1 ozaf048.142, https://doi.org/10.1093/mam/ozaf048.142

The White House (2022) OSTP issues guidance to make federally funded research freely available without delay. https://bidenwhitehouse.archives.gov/ostp/news-updates/2022/08/25/ostp-issues-guidance-to-make-federally-funded-research-freely-available-without-delay/

Wilkinson, M.D., Dumontier, M., Aalbersberg, I.J., Appleton, G., Axton, M., Baak, A., Blomberg, N., Boiten, J.-W., da Silva Santos, L.B., Bourne, P.E., Bouwman, J., Brookes, A.J., Clark, T., Crosas, M., Dillo, I., Dumon, O., Edmunds, S., Evelo, C.T., Finkers, R., Beltran, A.G., Gray, A.J.G., Groth, P., Goble, C., Grethe, J.S., Heringa, J., Hoen, P., Hooft, R., Kuhn, T., Kok, R., Kok, J., Lusher, S.J., Martone, M.E., Mons, A., Packer, A.L., Persson,

B., Rocca-Serra, P., Roos, M., van Schaik, R., Sansone, S.-A., Schultes, E., Sengstag, T., Slater, T., Strawn, G., Swertz, M.A., Thompson, M., van der Lei, J., van Mulligen, E., Velterop, J., Waagmeester, A., Wittenburg, P., Wolstencroft, K., Zhao, J., and Mons, B. (2016) The fair guiding principles for scientific data management and stewardship. *Scientific Data*, 3(1) 160018 https://doi-org/10.1038/sdata.2016.18

Wilkinson, S.R. Aloqalaa, M., Belhajjame, K.,Crusoe, M.R., de Paula Kinoshita, B., Gadelha, L., Garijo, D., Gustafsson, O.J.R., Juty, N., Kanwal, S., Khan, F.Z., Koster, J., Peters-von Gehlen, K., Pouchard, L., Rannow, R.K., Soiland-Reyes, S. Soranzo, N., Sufi, S., Sun, Z.,Vilne, B. Wouters, M.A., Yuen, D. & Carole Goble, C. (2025) Sci Data 12, 328 (2025). https://doi-org/10.1038/s41597-025-04451-9

Williams, C.A., Hyde, J., Smith, G.D.W., and Marquis, E.A. (2011) Effects of heavy-ion irradiation on solute segregation to dislocations in oxide-dispersion-strengthened Eurofer 97 steel. Journal of Nuclear Materials 412(1) 100-105 https://doi.org/10.1016/j.jnucmat.2011.02.029

Xu, Z., Li, D., Xu, W., Devaraj, A., Colby, R., Thevuthasan, S., Geiser, B.P., and Larson, D.J. (2015) Simulation of Heterogeneous Atom Probe Tip Shapes Evolution during Field Evaporation Using a Level Set Method and Different Evaporation Models. *Computer Physics Communications* 189 (April):106–13. https://doi.org/10.1016/j.cpc.2014.12.016

Yao, L., Withrow, T., Restrepo, O.D., Windl, W. and Marquis, E.A. (2015) Effects of the local structure dependence of evaporation fields on field evaporation behavior, *Applied Physics Letters*, 10724) 241602. https://doi.org/10.1063/1.4937454

Zanuttini, D., Blum, I., Rigutti, L., Vurpillot, F., Douady, J., Jacquet, E., Anglade, P.M., and Gervais, B. (2017) Simulation of field-induced molecular dissociation in atom-probe tomography: Identification of a neutral emission channel, Phys Rev A, 95, 6. https://doi.org/10.1103/PhysRevA.95.061401